\documentclass[twoside,english]{elsarticle}
\usepackage{ae,aecompl}
\usepackage[T1]{fontenc}
\usepackage[latin9]{inputenc}
\usepackage{geometry}
\usepackage{babel}
\usepackage{array}
\usepackage{float}
\usepackage{textcomp}
\usepackage{multirow}
\usepackage{amstext}
\usepackage{graphicx}
\usepackage[unicode=true]
 {hyperref}

\makeatletter

\newcommand*\LyXZeroWidthSpace{\hspace{0pt}}
\providecommand{\tabularnewline}{\\}
\newcommand{\lyxdot}{.}

\makeatletter
\def\ps@pprintTitle{%
  \let\@oddhead\@empty
  \let\@evenhead\@empty
  \def\@oddfoot{}
  \let\@evenfoot\@oddfoot
}
\makeatother

\usepackage{fancyhdr}
\usepackage{lineno}

\usepackage{pdflscape}
\usepackage[nolist,nohyperlinks]{acronym}

\makeatother

\begin{document}
\begin{frontmatter}

\title{{\Large{}Toward robust early-warning models:}\\
{\Large{}A horse race, ensembles and model uncertainty}\tnoteref{mytitlenote}}

\tnotetext[mytitlenote]{ We are grateful to Johannes Beutel, Andras Fulop, Benjamin Klaus,
Jan-Hannes Lang, Tuomas A. Peltonen, Roberto Savona, Gregor von Schweinitz,
Eero Tölö, Peter Welz and Marika Vezzoli for useful comments on previous
versions of the paper. The paper has also benefited from comments
during presentations at BITA'14 Seminar on Current Topics in Business,
IT and Analytics in Helsinki on 13 October 2014, seminars at the Financial
Stability Surveillance Division at the ECB in Frankfurt am Main on
21 November 2014 and 12 January 2015, a seminar at the Bank of Finland
in Helsinki on 28 November 2014, the 1st Conference on Recent Developments
in Financial Econometrics and Applications at Deakin University in
Geelong on 4--5 December 2014, the XXXVII Annual Meeting of the Finnish
Economic Association in Helsinki on 12 February 2015, 8th Financial
Risks International Forum on Scenarios, Stress and Forecasts in Finance
in Paris on 30--31 March 2015, seminar at the European Commission
Joint-Research Centre in Ispra on 13 April 2015, seminar at the University
of Brescia on 14 April 2015, Financial Stability seminar at the Deutsche
Bundesbank in Frankfurt am Main on 21 May 2015, the SYRTO conference
'A Critical Evaluation of Econometric Measures for Systemic Risk'
in Amsterdam on 5 June 2015, the INFINITI conference in Ljubljana,
Slovenia on 9 June 2015, Financial Stability seminar at the Bank of
Estonia on 30 June 2015, seminar at University of Pavia on 3 September
2015, keynote at the 5th CCCS Student Science Fair at University of
Basel on 16 September 2015, and seminar at Aalto University on 11
January 2016. The horse race in this paper is an implementation of
a proposal for a Lamfalussy Fellowship in 2012. Several of the methods
and exercises presented in this paper have also been implemented in
an online platform for interactive modeling (produced in conjunction
with and is property of infolytika): \href{http://cm.infolytika.com}{http://cm.infolytika.com}.
For further information see \citet{HolopainenCM2015}. The second
author thanks the GRI in Financial Services and the Louis Bachelier
Institute for financial support. All errors are our own. Corresponding
author: Peter Sarlin, Hanken School of Economics, Arkadiankatu 22,
00100 Helsinki, Finland, tel. +358405727670. E-mail: peter@risklab.fi.}

\author[P1]{Markus Holopainen}

\author[P2,P3,P1]{and Peter Sarlin}

\address[P2]{Center of Excellence SAFE at Goethe University Frankfurt, Germany}

\address[P3]{Department of Economics, Hanken School of Economics, Helsinki, Finland}

\address[P1]{RiskLab Finland at Arcada University of Applied Sciences, Helsinki,
Finland}
\begin{abstract}
This paper presents first steps toward robust models for crisis prediction.
We conduct a horse race of conventional statistical methods and more
recent machine learning methods as early-warning models. As individual
models are in the literature most often built in isolation of other
methods, the exercise is of high relevance for assessing the relative
performance of a wide variety of methods. Further, we test various
ensemble approaches to aggregating the information products of the
built models, providing a more robust basis for measuring country-level
vulnerabilities. Finally, we provide approaches to estimating model
uncertainty in early-warning exercises, particularly model performance
uncertainty and model output uncertainty. The approaches put forward
in this paper are shown with Europe as a playground. Generally, our
results show that the conventional statistical approaches are outperformed
by more advanced machine learning methods, such as $k$-nearest neighbors
and neural networks, and particularly by model aggregation approaches
through ensemble learning.\end{abstract}
\end{frontmatter}

\newpage{}

\section*{Non-technical summary}

The repeated occurrence of financial crises at the turn of the 21st
century has stimulated theoretical and empirical work on the phenomenon,
not least early-warning models. Yet, the history of these models goes
far back. Despite not always referring to macroprudential analysis,
the early days of risk analysis relied on assessing financial ratios
by hand rather than with advanced statistical methods on computers.
During the 1960s, discriminant analysis emerged, being the most dominantly
used technique until the 1980s. After the 1980s, DA has mainly been
replaced by logit/probit models. Applications of these models range
from early models for currency crises to recent ones on systemic financial
crises. In parallel, the simple yet intuitive signal extraction approach
that simply finds thresholds on individual indicators has gained popularity.
With technological advances, a soar in data availability and a thriving
need for progress in systemic risk identification, a new group of
flexible and non-linear machine learning techniques have been introduced
to various forms of financial stability surveillance. Recent literature
indicates that these novel approaches hold promise for systemic risk
identification because of their ability to identify and map complex
dependencies. The premise of difference in performance relates to
how methods treat two aspects: individual vs. multiple risk indicators
and linear vs. non-linear relationships. While the simplest approaches
linearly link individual indicators to crises, the more advanced techniques
account for both multiple indicators and different types of non-linearity,
such as the mapping of an indicator to crises and interaction effects
between multiple indicators.

Despite the fact that some methods hold promise over others, the use
and ranking of them is not an unproblematic task. This paper touches
upon three problem areas. First, there are few objective and thorough
comparisons of conventional and novel methods, and thus neither unanimity
on an overall ranking of methods nor on a single best-performing method.
Second, given an objective comparison, it is still unclear whether
one method can be generalized to outperform others on every single
dataset. It is not seldom that different approaches capture different
types of vulnerabilities, and hence can be seen to complement each
other. Despite potential differences in performance, this would contradict
the existence of one single best-in-class method, and instead suggest
value in simultaneous use of multiple approaches, or so-called ensembles.
Yet, the early-warning literature lacks a structured approach to the
use of multiple methods. Third, even if one could identify the best-performing
methods and come up with an approach to make use of multiple methods
simultaneously, the literature on early-warning models lacks measures
of statistical significance or uncertainty. Although crisis probabilities
may breach a threshold, there is no work testing the possibility of
an exceedance to have occurred due to sampling error alone. Likewise,
little or no attention has been given to testing equality of two methods'
early-warning performance or individual probabilities and thresholds.

This paper aims at providing a solution to all of the three above
mentioned challenges. First, we conduct an objective horse race of
methods for early-warning models, including a large number of common
techniques from conventional statistics and machine learning, with
a particular focus on the problem as a classification task. The objectivity
of the exercise derives from identical sampling into in-sample and
out-of-sample data for each method, identical model selection, and
identical model specification. For generalizability and comparability,
we make use of cross-validation and recursive real-time estimation
to assure that and assess how results generalize to out-of-sample
data. The two exercises differ in their sampling of data, particularly
the in-sample and out-of-sample partitions used for each estimation.
While cross-validation is common in machine learning and allows an
efficient use of small samples, exercises may benefit from the fact
that data are sampled randomly despite most likely exhibiting time
dependence. The recursive exercises, on the contrary, account for
time dependence in data by strictly using historical samples for out-of-sample
predictions, which nevertheless requires more data, particularly in
the time-series dimension. These two exercises allow exploring performance
across methods, and how that is impacted by the evaluation exercise.

Second, acknowledging the fact that no one method can be generalized
to outperform all others, we put forward two strands of approaches
for the simultaneous use of multiple methods. A natural starting point
is to collect model signals from all methods in the horse race, in
order to assess the number of methods that signal for a given country
at a given point in time. Two structured approaches involve choosing
the best method (in-sample) for out-of-sample use, and relying on
the majority vote of all methods together. Then, moving toward more
standard ensemble methods for the use of multiple methods, we combine
model output probabilities into an arithmetic mean of all methods.
With potential further gains in aggregation, we take a performance-weighted
mean by letting methods with better in-sample performance contribute
more to the aggregated model output. Third, we provide approaches
to testing statistical significance in early-warning exercises, including
both model performance and output uncertainty. With the sampling techniques
of repeated cross-validation and bootstrapping, we estimate properties
of the performance of models, and may hence test for statistical significance
when ranking models. Further, through sampling techniques, we may
also use the variation in model output and thresholds to compute properties
for capturing their reliability for individual observations. Beyond
confidence bands for representation of uncertainty, this also provides
a basis for hypothesis testing, in which an interest of importance
ought to be whether a model output is statistically significantly
different from the cut-off threshold.

The approaches put forward in this paper are illustrated in a European
setting, for which we use a large number of macro-financial indicators
for 15 European economies since the 1980s. First, we present rankings
of all methods for the objective horse race, after which we proceed
to aggregation and statistical significance tests. Generally, our
results show that the classical approaches are outperformed by more
advanced machine learning methods, such as $k$-nearest neighbors
and neural networks, in terms of the Usefulness and Area Under the
Curve (AUC) measures. This holds for both horse race exercises. While
several of the differences in rankings are statistically insignificant,
a particular finding is the outperformance of ensemble models, which
is significant in both exercises. More importantly, the objective
exercises in this paper provide strong evidence that early-warning
modeling in general is a useful tool to identify systemic risk at
an early stage.

\section*{\newpage{}}

\section{Introduction}

Systemic risk measurement lies at the very core of macroprudential
oversight, yet anticipating financial crises and issuing early warnings
is intrinsically difficult. The literature on early-warning models
has, nevertheless, shown that it is no impossible task. This paper
provides a three-fold contribution to the early-warning literature:
(\emph{i}) a horse race of early-warning methods, (\emph{ii}) approaches
to aggregating model output from multiple methods, and (\emph{iii})
model performance and output uncertainty.

The repeated occurrence of financial crises at the turn of the 21st
century has stimulated theoretical and empirical work on the phenomenon,
not least early-warning models. Yet, the history of these models goes
far back. Despite not always referring to macroprudential analysis,
the early days of risk analysis relied on assessing financial ratios
by hand rather than with advanced statistical methods on computers
(e.g., \citet{RamserFoster1931}). After Beaver\textquoteright s \citep{Beaver1966}
seminal work on a univariate approach to discriminant analysis (DA),
\citet{Altman1968} further developed DA for multivariate analysis.
Even though DA suffers from frequently violated assumptions like normality
of the indicators, it was the dominant technique until the 1980s.
\citet{FrankJr1971327} and \citet{TAffler1984}, for example, used
DA for predicting sovereign debt crises. After the 1980s, DA has mainly
been replaced by logit/probit models. Applications of these models
range from the early model for currency crises by \citet{Frankel1996}
to a recent one on systemic financial crises by \citet{Duca2012}.
In parallel, the simple yet intuitive signal extraction approach that
simply finds thresholds on individual indicators has gained popularity,
again ranging from early work on currency crises by \citet{Kaminskyetal1998b}
to later work on costly asset booms by \citet{Alessi2011520}. Yet,
these methods suffer from assumptions violated more often than not,
such as fixed distributional relationship between the indicators and
the response (e.g., logistic/normal), and the absence of interactions
between indicators (e.g., non-linearities in crisis probabilities
with increases in fragilities). With technological advances, a soar
in data availability and a thriving need for progress in systemic
risk identification, a new group of flexible and non-linear machine
learning techniques have been introduced to various forms of financial
stability surveillance. Recent literature indicates that these novel
approaches hold promise for systemic risk identification (e.g., as
reviewed in \citet{DemyanykHasan2010} and \citet{Sarlin2013NG}).\footnote{See also a number of applications, such as \citet{NagMitra1999},
\citet{FranckSchmied2003}, \citet{Peltonen2006}, \citet{SarlinMarghescu2011a},
\citet{SarlinPeltonen2013,Sarlin2013a} and \citet{AlessiDetken2014}.} The premise of difference in performance relates to how methods treat
two aspects: individual vs. multiple risk indicators and linear vs.
non-linear relationships. While the simplest approaches linearly link
individual indicators to crises, the more advanced techniques account
for both multiple indicators and different types of non-linearity,
such as the mapping of an indicator to crises and interaction effects
between multiple indicators.

Despite the fact that some methods hold promise over others, the use
and ranking of them is not an unproblematic task. This paper touches
upon three problem areas. First, there are few objective and thorough
comparisons of conventional and novel methods, and thus neither unanimity
on an overall ranking of methods nor on a single best-performing method.
Though the horse race conducted among members of the Macro-prudential
Research Network of the European System of Central Banks aims at a
prediction competition, it does not provide a solid basis for objective
performance comparisons \citep{Alessietal2014}. Even though disseminating
information of models underlying discretionary policy discussion is
a valuable task, the panel of presented methods are built and applied
in varying contexts. This relates more to a horse show than a horse
race. Second, given an objective comparison, it is still unclear whether
one method can be generalized to outperform others on every single
dataset. It is not seldom that different approaches capture different
types of vulnerabilities, and hence can be seen to complement each
other. Despite potential differences in performance, this would contradict
the existence of one single best-in-class method, and instead suggest
value in simultaneous use of multiple approaches, or so-called ensembles.
Yet, the early-warning literature lacks a structured approach to the
use of multiple methods. Third, even if one could identify the best-performing
methods and come up with an approach to make use of multiple methods
simultaneously, the literature on early-warning models lacks measures
of statistical significance or uncertainty. Moving beyond the seminal
work by \citet{El-shagietal2013}, where the authors put forward approaches
for assessing the null of whether or not a model is useful, there
is a lack of work estimating statistically significant differences
in performance among methods. Likewise, although crisis probabilities
may breach a threshold, there is no work testing the possibility of
an exceedance to have occurred due to sampling error alone. While
\citet{Hurlinetal2013} provide a general-purpose equality test for
firms' risk measures, little or no attention has been given to testing
equality of two methods' early-warning performance or individual probabilities
and thresholds.

This paper aims at providing a solution to all of the three above
mentioned challenges. First, we conduct an objective horse race of
methods for early-warning models, including a large number of common
techniques from conventional statistics and machine learning, with
a particular focus on the problem as a classification task. The objectivity
of the exercise derives from identical sampling into in-sample and
out-of-sample data for each method, identical model selection, and
identical model specification. For generalizability and comparability,
we make use of cross-validation and recursive real-time estimation
to assure that and assess how results generalize to out-of-sample
data. Rather than an absolute ranking that could be generalized to
any context, this provides evidence on the potential in more advanced
machine learning approaches in these types of exercises, as well as
points to the importance of using appropriate resampling techniques,
such as accounting for time dependence. Second, acknowledging the
fact that no one method can be generalized to outperform all others,
we put forward two strands of approaches for the simultaneous use
of multiple methods. A natural starting point is to collect model
signals from all methods in the horse race, in order to assess the
number of methods that signal for a given country at a given point
in time. Two structured approaches involve choosing the best method
(in-sample) for out-of-sample use, and relying on the majority vote
of all methods together. Then, moving toward more standard ensemble
methods for the use of multiple methods, we combine model output probabilities
into an arithmetic mean of all methods. With potential further gains
in aggregation, we take a performance-weighted mean by letting methods
with better in-sample performance contribute more to the aggregated
model output. Third, we provide approaches to testing statistical
significance in early-warning exercises, including both model performance
and output uncertainty. With the sampling techniques of repeated cross-validation
and bootstrapping, we estimate properties of the performance of models,
and may hence test for statistical significance when ranking models.
Further, through sampling techniques, we may also use the variation
in model output and thresholds to compute properties for capturing
their reliability for individual observations. Beyond confidence bands
for representation of uncertainty, this also provides a basis for
hypothesis testing, in which an interest of importance ought to be
whether a model output is statistically significantly different from
the cut-off threshold. 

The approaches put forward in this paper are illustrated in a European
setting, for which we use a large number of macro-financial indicators
for 15 European economies since the 1980s. First, we present rankings
of all methods for the objective horse race, after which we proceed
to aggregation and statistical significance tests. Generally, our
results show that the classical approaches are outperformed by more
advanced machine learning methods, such as $k$-nearest neighbors
and neural networks, in terms of the Usefulness and Area Under the
Curve (AUC) measures. This holds for both horse race exercises. While
several of the differences in rankings are statistically insignificant,
a particular finding is the outperformance of ensemble models, which
is significant in both exercises. More importantly, the objective
exercises in this paper provide strong evidence that early-warning
modeling in general is a useful tool to identify systemic risk at
an early stage.

This paper is organized as follows. In Section 2, we describe the
used data, including indicators and events, the methods for the early-warning
models, and estimation strategies. Then, we present the set-up for
the horse race, as well as approaches for aggregating model output
and computing model uncertainty. In Section 4, we present results
of the horse race, its aggregations, and model uncertainty in a European
setting. Finally, we conclude in Section 5.

\section{Data and methods}

This section presents the data and methods used in the paper. Whereas
the dataset covers both crisis event definitions and vulnerability
indicators, the methods include classification techniques ranging
from conventional statistical modeling to more recent machine learning
algorithms.

\subsection{Data}

The dataset used in this paper has been collected with the aim of
covering as many European economies as possible. While a focus on
similar economies might improve homogeneity in early-warning models,
we aim at collecting a dataset as large as possible for the data-demanding
estimations. The data used in this paper are quarterly and span from
1976Q1 to 2014Q3. The sample is an unbalanced panel with 15 European
Union countries: Austria, Belgium, Denmark, Finland, France, Germany,
Greece, Ireland, Italy, Luxembourg, the Netherlands, Portugal, Spain,
Sweden, and the United Kingdom. In total, the sample includes 15 crisis
events, which cover systemic banking crises. The dataset consists
of two parts: crisis events and vulnerability indicators. In the following,
we provide a more detailed description of the two parts.

\paragraph{Crisis events}

The crisis events used in this paper are chosen as to cover country-level
distress in the financial sector. We are concerned with banking crises
with systemic implications and hence mainly rely on the IMF's crisis
event initiative by \citet{LaevenValencia2013}. Yet, as their database
is partly annual, we complement our events with starting dates from
the quarterly database collected by the European System of Central
Banks (ESCB) Heads of Research Group, and as reported in \citet{Babeckyetal2012}.
The database includes banking, currency and debt crisis events for
a global set of advanced economies from 1970 to 2012, of which we
only use systemic banking crisis events.\footnote{To include events after 2012, as well as some minor amendments to
the original event database by Babecky et al. (2013), we rely on an
update by the Countercyclical Capital Buffer Working Group within
the ESCB.} In general, both of the above databases are a compilation of crisis
events from a large number of influential papers, which have been
complemented and cross-checked by ESCB Heads of Research. The paper
with which the events have been cross-checked include \citet{KindlebergerAliber2011},
\citet{IMF2010}, \citet{ReinhartRogoff2009}, \citet{CaprioKlingebiel2003},
\citet{Caprioetal2005}, and \citet{Kaminsky1996} among many others.

\paragraph{Early-warning indicators}

The second part of the dataset consists of a number of country-level
vulnerability indicators. Generally, these cover a range of macro-financial
imbalances. We include measures covering asset prices (e.g., house
and stock prices), leverage (e.g., mortgages, private loans and household
loans), business cycle indicators (GDP and inflation), measures from
the EU Macroeconomic Imbalance Procedure (e.g., current account deficits
and government debt), and the banking sector (e.g., loans to deposits).
In most cases, we have relied on the most commonly used transformation,
such as ratios to GDP or income, growth rates, and absolute and relative
deviations from a trend. The indicators are sourced from Eurostat,
OECD, ECB Statistical Data Warehouse and the BIS Statistics. 

For detrending, the trend is extracted using one-sided Hodrick\textendash Prescott
filter (HP filter). This means that each point of the trend line corresponds
to the ordinary HP trend calculated recursively from the beginning
of the series to each point in time. By doing this, we do not use
future information when calculating the trend, but rather use the
information set available to the policymaker at each point in time.
The smoothness parameter of the HP filter is specified to be 400 000
as suggested by \citet{Drehmann2011}. This has been suggested to
appropriately capture the nature of financial cycles in quarterly
data. Growth rates are defined to be annual, whereas we follow \citet{laina2015}
by using both absolute and relative deviations from trend, of which
the latter differs from the former by relating the deviation to the
value of the trend. The indicators used in this paper combine several
sources for broad coverage and for deriving ratios of appropriate
variables, and are presented in Table \ref{tab:Indicators}. Their
descriptive statistics are shown in Table \ref{tab:Indicators-descriptivestatistics}. 

As proper use of data is essential in order to obtain an objective
indication of the usefulness of any modeling approach, a note regarding
the relationship between crisis events and indicators is in order.
Whilst the uncertainty regarding the definitions of crisis events
cannot be disputed, this holds true for any empirical exercise. To
visualize the relationship between the actual crisis events and the
indicators, as well as their lead time, we include time-series plots
for each indicator from $t-12$ to $t+8$ around crisis occurrences
in Figure \ref{fig:Time-series-plots-of-indicators}. The figure illustrates
that patterns of several indicators, such as the credit gap and asset
price changes, for instance, take elevated values prior to crisis
events, which is indeed in line with the early-warning literature.

\begin{table}[H]
\protect\caption{\label{tab:Indicators}A list of indicators.}

\noindent \centering{}\includegraphics[width=1\textwidth]{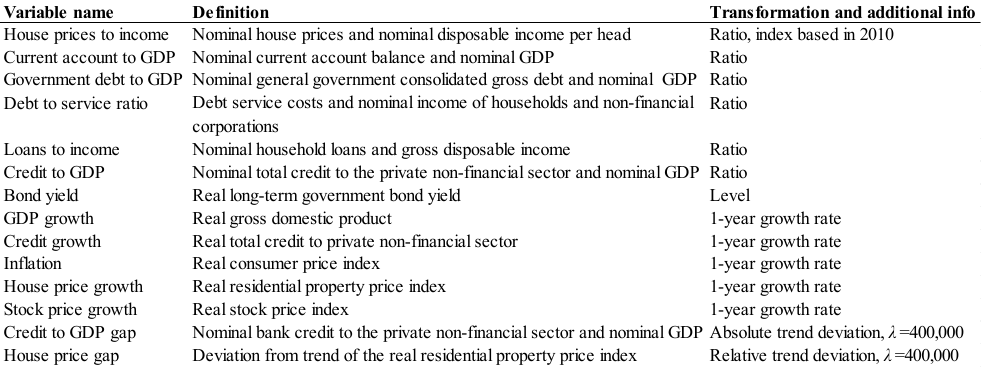}
\end{table}

\begin{table}[H]
\protect\caption{\label{tab:Indicators-descriptivestatistics}Descriptive statistics
of indicators.}

\noindent \centering{}\includegraphics[width=0.8\textwidth]{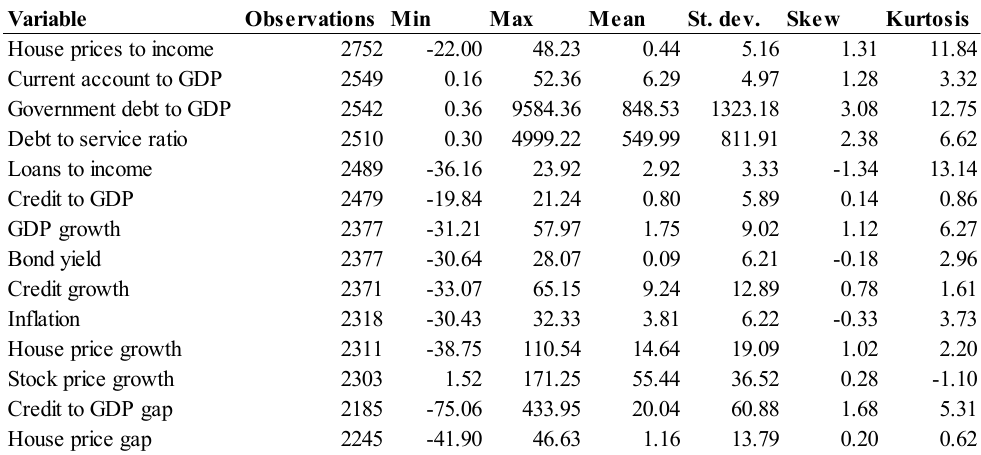}
\end{table}

\subsection{Early warning as a classification problem\label{sub:Early-warning-as}}

Early-warning models require evaluation criteria that account for
the nature of the underlying problem, which relates to low-probability,
high-impact events. It is of central importance that the evaluation
framework resembles the decision problem faced by a policymaker. The
signal evaluation framework focuses on a policymaker with relative
preferences between type I and II errors, and the usefulness that
she derives by using a model, in relation to not using it. In the
vein of the loss-function approach proposed by \citet{Alessi2011520},
the framework applied here follows an updated and extended version
in \citet{Sarlin2013b}.

To mimic an ideal leading indicator, we build a binary state variable
$C_{n}(h)\in\left\{ 0,1\right\} $ for observation $n$ (where $n=1,2,\ldots,N$)
given a specified forecast horizon $h$. Let $C_{n}(h)$ be a binary
indicator that is one during pre-crisis periods and zero otherwise.
For detecting events $C_{n}$ using information from indicators, we
need to estimate the probability of being in a vulnerable state $p_{n}\in\left[0,1\right]$.
Herein, we make use of a number of different methods $m$ for estimating
$p_{n}^{m}$, ranging from the simple signal extraction approach to
more sophisticated techniques from machine learning. The probability
$p_{n}$ is turned into a binary prediction $B_{n}$, which takes
the value one if $p_{n}$ exceeds a specified threshold $\tau\in\left[0,1\right]$
and zero otherwise. The correspondence between the prediction $B_{n}$
and the ideal leading indicator $C_{n}$ can then be summarized into
a so-called contingency matrix, as described in Table \ref{tab:A-contingency-matrix.}. 

\begin{table}[H]
\protect\caption{\label{tab:A-contingency-matrix.}A contingency matrix.}

\noindent \centering{}%
\begin{tabular}{|c|c|c|c|}
\cline{3-4} 
\multicolumn{1}{c}{} &  & \multicolumn{2}{c|}{{\small{}Actual class $C_{n}$}}\tabularnewline
\cline{3-4} 
\multicolumn{1}{c}{} &  & {\small{}Pre-crisis period} & {\small{}Tranquil period}\tabularnewline
\hline 
\multirow{4}{*}{{\small{}Predicted class $P_{n}$}} & \multirow{2}{*}{{\small{}Signal}} & {\small{}Correct call} & {\small{}False alarm}\tabularnewline
 &  & \emph{\small{}True positive (TP)} & \emph{\small{}False positive (FP)}\tabularnewline
\cline{2-4} 
 & \multirow{2}{*}{{\small{}No signal}} & {\small{}Missed crisis} & {\small{}Correct silence}\tabularnewline
 &  & \emph{\small{}False negative (FN)} & \emph{\small{}True negative (TN)}\tabularnewline
\hline 
\end{tabular}
\end{table}

The frequencies of prediction-realization combinations in the contingency
matrix can be used for computing measures of classification performance.
A policymaker can be thought to be primarily concerned with two types
of errors: issuing a false alarm and missing a crisis. The evaluation
framework described below is based upon that in \citet{Sarlin2013b}
for turning policymakers' preferences into a loss function, where
the policymaker has relative preferences between type I and II errors.
While type I errors represent the share of missed crises to the frequency
of crises $T_{1}\in\left[0,1\right]=$FN/(TP+FN), type II errors represent
the share of issued false alarms to the frequency of tranquil periods
$T_{2}\in\left[0,1\right]=$FP/(FP+TN). Given probabilities $p_{n}$
of a model, the policymaker then finds an optimal threshold $\tau^{*}$
such that her loss is minimized. The loss of a policymaker includes
$T_{1}$ and $T_{2}$, weighted by relative preferences between missing
crises ($\mu$) and issuing false alarms ($1-\mu$). By accounting
for unconditional probabilities of crises $P_{1}=\mbox{Pr}(C=1)$
and tranquil periods $P_{2}=\mbox{Pr}(C=0)=1-P_{1}$, as classes are
not of equal size and errors are scaled with class size, the loss
function can be written as follows:

\begin{equation}
L(\mu)=\mu T_{1}P_{1}+(1-\mu)T_{2}P_{2}.
\end{equation}
Further, the Usefulness of a model can be defined in a more intuitive
manner. First, the absolute Usefulness ($U_{a}$) is given by: 

\begin{equation}
U_{a}(\mu)=\mbox{min}(\mu P_{1},\left(1-\mu\right)P_{2})-L(\mu),
\end{equation}
which computes the superiority of a model in relation to not using
any model. As the unconditional probabilities are commonly unbalanced
and the policymaker may be more concerned about the rare class, a
policymaker could achieve a loss of $\mbox{min}(\mu P_{1},\left(1-\mu\right)P_{2})$
by either always or never signaling a crisis. This predicament highlights
the challenge in building a Useful early-warning model: With a non-perfect
model, it would otherwise easily pay-off for the policymaker to always
signal the high-frequency class. Second, we can compute the relative
Usefulness $U_{r}$ as follows: 

\begin{equation}
U_{r}(\mu)=\frac{U_{a}(\mu)}{\mbox{min}(\mu P_{1},\left(1-\mu\right)P_{2})},
\end{equation}
where $U_{a}$ of the model is compared with the maximum possible
usefulness of the model. That is, the loss of disregarding the model
is the maximum available Usefulness. Hence, $U_{r}$ reports $U_{a}$
as a share of the Usefulness that a policymaker would gain with a
perfectly-performing model, which supports interpretation of the measure.
It is worth noting that $U_{a}$ better lends to comparisons over
different $\mu$.

Beyond the above measures, the contingency matrix may be used for
computing a wide range of other quantitative measures.\footnote{Some of the commonly used evaluation measures include: Recall positives
(or TP rate) = TP/(TP+FN), Recall negatives (or TN rate) = TN/(TN+FP),
Precision positives = TP/(TP+FP), Precision negatives = TN/(TN+FN),
Accuracy = (TP+TN)/(TP+TN+FP+FN), FP rate = FP/(FP+TN), and FN rate
= FN/(FN+TP)} Receiver operating characteristics (ROC) curves and the area under
the ROC curve (AUC) are also used for comparing performance of early-warning
models and indicators. The ROC curve plots, for the complete range
of $\tau\in[0,1]$, the conditional probability of positives to the
conditional probability of negatives: 

\[
ROC=\frac{\mbox{Pr}(P=1\mid C=1)}{1-\mbox{Pr}(P=0\mid C=0)}.
\]

\subsection{Classification methods}

The purpose of any classification algorithm is to identify to which
of a set of classes a new observation belongs, based on one or more
predictor variables. Classification is considered an instance of supervised
learning, where a training set of correctly identified observations
is available. In this paper, a number of probabilistic classifiers
are used, whose outputs are probabilities indicating to which of the
qualitative classes an observation belongs. In our case, the dependent
(or outcome) variable represents the two classes of pre-crisis periods
(1) and tranquil periods (0).

Generally, a classifier attempts to assign each observation to the
most likely class, given its predictor values. For the binary case,
where there are only two possible classes, an optimal classifier (which
minimizes the error rate) predicts class one if $\mbox{Pr}(Y=1|X=x)>0.5$,
and class zero otherwise. This classifier is denoted as the Bayes
classifier. Ideally, one would like to predict qualitative responses
using the Bayes classifier, but for real-world data, however, the
conditional distribution of $Y$ given $X$ is unknown. Thus, the
goal of many approaches is to estimate this conditional distribution
and classify an observation to the category with the highest estimated
probability. For real-world applications, it may also be noted that
the optimal threshold $\tau$ between classes is not always 0.5, but
varies. This optimal threshold may be a result of optimizing the above
discussed Usefulness, and is examined in further detail later in the
paper.

\begin{figure}[H]
\begin{centering}
\includegraphics[width=0.9\textwidth]{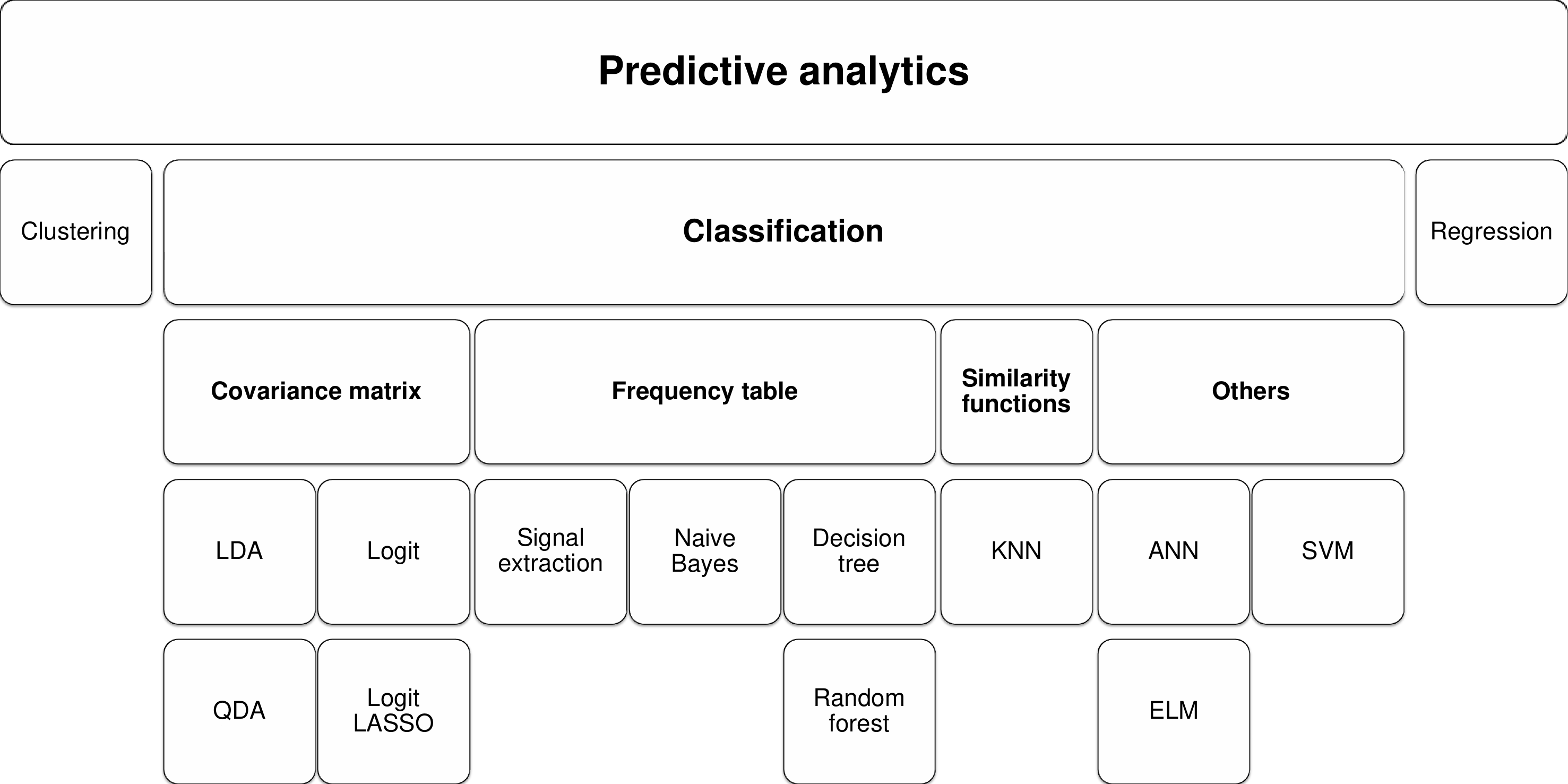}
\par\end{centering}

\protect\caption{\label{fig:A-taxonomy-of-methods}A taxonomy of classification methods}
\end{figure}

This paper aims to gather a versatile set of different classification
methods, from the simple approach of signal extraction to the considerably
more computationally intensive neural networks and support vector
machines. The methods used for deriving early-warning models have
been put into context in Figure \ref{fig:A-taxonomy-of-methods} and
papers applying these methods in an early-warning exercise have been
reviewed in Table \ref{tab:Literature-review}. The methods are presented
in more detail below.

\paragraph{Signal extraction}

The signal extraction approach introduced by \citet{Kaminskyetal1998b}
simply analyzes the level of an indicator, and issues a signal if
the value exceeds a specified threshold. In order to issue binary
signals, we specify the threshold value as to optimize classification
performance, which is herein measured with relative Usefulness \citep{Kaminsky1996}.
However, the key limitation of this approach is that it does not enable
any interaction between or weighting of indicators, while an advantage
is that it demonstrates a more direct measure of the importance and
provides a ranking of each indicator.\footnote{We are aware of the multivariate signal extraction, but do not consider
it herein as we judge logit analysis, among others, to cover the idea
of estimating weights for transforming multiple indicators into one
output.} Despite this, it is one of the most commonly applied early-warning
techniques.

\paragraph{Linear Discriminant Analysis (LDA)}

LDA, introduced by \citet{Fisher1936}, is a commonly used method
in statistics for expressing one dependent variable as a linear combination
of one or more continuous predictors. LDA assumes that the predictor
variables are normally distributed, with a mean vector and a common
covariance matrix for all classes, and implements Bayes' theorem to
approximate the Bayes classifier. LDA has been shown to perform well
on small data sets, if the above-mentioned conditions apply. Yet even
though DA suffers from the frequently violated assumptions, it was
the dominant technique until the 1980s, after which it was oftentimes
replaced by logit/probit models.

\paragraph{Quadratic Discriminant Analysis (QDA)}

QDA is a variant of LDA, which estimates a separate covariance matrix
for each class (see, e.g., \citet{VenablesRipley2002}). This causes
the number of parameters to estimate to rise significantly, but consequently
results in a non-linear decision boundary. To the best of our knowledge,
QDA has not been applied for early-warning exercises at the country
level.

\paragraph{Logit analysis}

Much of the early-warning literature deals with models that rely on
logit/probit regression. Logit analysis uses the logistic function
to describe the probability of an observation belonging to one of
two classes, based on a regression of one or more continuous predictors.
For the case with one predictor variable, the logistic function is
$p(X)=\frac{e^{\beta_{0}+\beta_{1}X}}{1+e^{\beta_{0}+\beta_{1}X}}$.
From this, it is obvious to extend the function to the case of several
predictors. Logit and probit models have frequently been applied to
predicting financial crises, as can be seen in an early review by
\citet{Bergetal2005}. However, the distributional (logistic/normal)
assumption on the relationship between the indicators and the response
as well as the absence of interactions between variables may often
be violated. \citet{Duca2012}, for example, show that the probability
of a crisis increases non-linearly as the number of fragilities increase.

\paragraph{Logit LASSO}

The LASSO (Least Absolute Shrinkage and Selection Operator) logistic
regression (\citet{Tibshirani1996}) attempts to select the most relevant
predictor variables for inference and is often applied to problems
with a large number of predictors. The method maximizes the log likelihood
subject to a bound on the sum of absolute values of the coefficients
$\mathrm{max}{}_{\beta}l(\beta\mid y)-\lambda\sum_{i}\mid\beta_{i}\mid$,
for which the $\mid\beta_{i}\mid$ is penalized by the $L_{1}$ norm.
This implies that the LASSO sets some coefficients to equal zero,
and produces sparse models with a simultaneous variable selection.
The optimal penalization parameter $\lambda$ is oftentimes chosen
empirically via cross-validation. We are only aware of the use of
the Logit LASSO in this context in \citet{Langetal2015}, wherein
it is mainly used to identify risks in bank-level data, but also aggregated
to the country level for assessing risks in entire banking sectors.

\paragraph{Naive Bayes}

In machine learning, the Naive Bayes method is one of the most common
Bayesian network methods (see e.g. \citet{Kohavietal1997}). Bayesian
learning is based on calculating the probability of each hypothesis
(or relation between predictor and response), given the data. The
method is called 'naive' as it assumes that the predictor variables
are conditionally independent. Consequently, the method may give high
weights to several predictors which are correlated, unlike the methods
discussed above, which balance the influence of all predictors. However,
the method has been known to scale well to large problems. To the
best of our knowledge, Naive Bayes has not been applied for early-warning
exercises at the country level.

\paragraph{$k$-nearest neighbors (KNN)}

KNN is a non-parametric method which uses similarity functions to
determine the class of an observation based on its $k$ nearest observations
(see, e.g. \citet{Altman1992}). Given a positive integer $k$ and
an observation $x_{0}$, the algorithm first identifies the $k$ points
$x_{k}$ in the data closest to $x_{0}$. The probability for belonging
to a class is then estimated as the fraction of the $k$ closest points,
whose response values correspond with the respective class. The method
is considered to be among the simplest in the realm of machine learning,
and has two free parameters, the integer $k$ and a parameter which
affects the search distance for neighbors, which can be optimized
for each data set. As with Naive Bayes, we are not aware of previous
use of KNN in early-warning exercises at the country level.

\paragraph{Classification trees}

Classification trees, as discussed by \citet{Breimanetal1984}, implement
a decision tree-type structure, which reach a decision by performing
a sequence of tests on the values of the predictors. In a classification
tree, the classes are represented by leaves, and the conjunctions
of predictors are represented by the branches leading to the classes.
These conjunction rules segment the predictor space into a number
of simpler regions, allowing for decision boundaries of complex shapes.
Given similar loss functions, an identical result could also be reached
through sequential signal extraction. The method has proven successful
in many areas of machine learning, and has the advantage of high interpretability.
To reduce complexity and improve generalizability, sections of the
tree are often pruned until optimal out-of-sample performance is reached.
The degree of pruning is determined by a complexity parameter, which
is used in this paper as a free parameter. In the early-warning literature,
the use of classification trees has been fairly common.

\paragraph{Random forest}

The random forest method, introduced by \citet{Breiman2001}, uses
classification trees as building blocks to construct a more sophisticated
method, at the expense of interpretability. The method grows a number
of classification trees based on differently sampled subsets of the
data. Additionally, at each split, a randomly selected sample is drawn
from the full set of predictors. Only predictors from this sample
are considered as candidates for the split, effectively forcing diversity
in each tree. Lastly, the average of all trees is calculated. As there
is less correlation between the trees, this leads to a reduction in
variance in the average. In this paper, two free parameters are considered:
the number of trees, and the number of predictors sampled as candidates
at each split. To the best of our knowledge, random forests have only
been applied to early-warning exercises in \citet{AlessiDetken2014}.

\paragraph{Artificial Neural Networks (ANN)}

Inspired by the functioning of neurons in the human brain, ANNs are
composed of nodes or units connected by weighted links (see, e.g.,
\citet{VenablesRipley2002}). These weights act as network parameters
that are tuned iteratively by a learning algorithm. The simplest type
of ANN is the single hidden layer feed-forward neural network (SLFN),
which has one input, hidden and output layer. The input layer distributes
the input values to the units in the hidden layer, whereas the unit(s)
in the output layer computes the weighted sum of the inputs from the
hidden layer, in order to yield a classifier probability. Despite
ANNs with no size restrictions are universal approximators for any
continuous function \citep{Horniketal1989}, computation time increases
exponentially and their interpretability diminishes as ANNs grow in
size. Further, discriminant and logit/probit analysis can in fact
be related to very simple ANNs \citep{Ripley1994,Sarle1994}: so-called
single-layer perceptrons (i.e., no hidden layer) with a threshold
and logistic activation function. This paper uses a basic SLFN with
three free parameters: the number of units in the hidden layer, the
maximum number of iterations, and the weight decay. The first parameter
controls the complexity of the network, while the last two are used
to control how the learning algorithm converges. The use of ANNs has
been fairly common in the academic early-warning literature.

\paragraph{Extreme Learning Machines (ELM)}

As introduced by \citet{Huangetal2006}, the ELM refers to a specific
learning algorithm used to train a SLFN-type neural network. Unlike
conventional iterative learning algorithms, the ELM algorithm randomizes
the input weights and analytically determines the output weights of
the network. When trained with this algorithm, the SLFN generally
requires a higher number of units in the hidden layer, but computation
time is greatly reduced and the resulting neural network may have
better generalization ability. In this paper, two free parameters
are considered: the number of units in the hidden layer, and the type
of activation function used in the network. To the best of our knowledge,
we are not aware of previous applications of the ELM algorithm to
crisis prediction.

\paragraph{Support Vector Machines (SVM)}

The SVM, introduced by \citet{CortesVapnik1995}, is one of the most
popular machine learning methods for supervised learning. It is a
non-parametric method that uses hyperplanes in a high-dimensional
space to construct a decision boundary for a separation between classes.
It comes with several desirable properties. First, an SVM constructs
a maximum margin separator, i.e. the chosen decision boundary is the
one with the largest possible distance to the training data points,
enhancing generalization performance. Second, it relies on support
vectors when constructing this separator, and not on all the data
points, such as in logistic regression. These properties lead to the
method having high flexibility, but still being somewhat resistant
to overfitting. However, SVMs lack interpretability. The free parameters
considered are: the cost parameter, which affects the tolerance for
misclassified observations when constructing the separator; the gamma
parameter, defining the area of influence for a support vector; and
the kernel type used. We are not aware of studies using SVMs for the
purpose of deriving early-warning models.

\begin{table}[H]
\centering{}\protect\caption{\label{tab:Literature-review}Literature review.}
{\footnotesize{}}%
\begin{tabular}{cccc}
{\footnotesize{}Method} & Currency crisis & Sovereign crisis & Banking crisis\tabularnewline
\hline 
\hline 
{\footnotesize{}Signal extraction} & \citep{Kaminskyetal1998b} & \citep{KnedlikSchweinitz2012} & \citep{BorioDrehman2009} \citep{Alessi2011520}\tabularnewline
{\footnotesize{}LDA} & -- & \citep{FrankJr1971327} \citep{TAffler1984} & --\tabularnewline
{\footnotesize{}QDA} & -- & -- & --\tabularnewline
{\footnotesize{}Logit} & \citep{Eichengreen1998} \citep{Frankel1996} \citep{Sachs1996} \citep{Berg1999a}
\citep{Bussiere2006953}  & \citep{Schmidt1984} \citep{FuertesKalotychou2006} & \citep{Barreletal2010} \citep{Duca2012}\tabularnewline
{\footnotesize{}Logit LASSO} & -- & -- & \citep{Langetal2015}\tabularnewline
{\footnotesize{}KNN} & -- & -- & --\tabularnewline
{\footnotesize{}Trees} & \citep{Kaminksky2003} \citep{Chamonetal2007} & \citep{Schimmelpfennigetal2003} & \citep{Duttaguptaetal2011}\tabularnewline
{\footnotesize{}Random forest} & -- & -- & \citep{AlessiDetken2014}\tabularnewline
{\footnotesize{}ANN} & \citep{NagMitra1999} \citep{FranckSchmied2003} \citep{Peltonen2006}
\citep{SarlinMarghescu2011a} & \citep{Fioramanti2008} & \citep{Sarlin2013NG}\tabularnewline
{\footnotesize{}ELM} & -- & -- & --\tabularnewline
{\footnotesize{}SVM} & -- & -- & --\tabularnewline
\end{tabular}{\footnotesize \par}
\end{table}

\section{Horse race, aggregation and model uncertainty}

This section presents the methodology behind the robust and objective
horse race and its aggregation, as well as approaches for estimating
model uncertainty.

\subsection{Set-up of the horse race\label{sub:Set-up-of-the-horse-race}}

To continue from the data, classification problem and methods presented
in Section 2, we herein focus on the set-up for and parameters used
in the horse race, ranging from details in the use of data and general
specification of the classification problem to estimation strategies
and modeling. The aim of the set-up is to mimic real-time use as much
as possible by both using data in a realistic manner and tackling
the classification problem using state-of-the-art specifications.
The specification needs also to be generic in nature, as the objectivity
of a horse race relies on applying the same procedures to all methods.

\paragraph{Model specifications}

This section describes the choices regarding model specifications
that underlie the exercises in this paper. In all choices, we have
tried to follow the convention in the most recent literature on the
topic. Despite the fact that model output is country-specific, the
literature has preferred the use of pooled data and models (e.g.,
\citet{FuertesKalotychou2006,SarlinPeltonen2013}). In theory, one
would desire to account for country-specific effects describing crises,
but the rationale behind pooled models descends from the aim to capture
a wide variety of crises and the relatively small number of events
in individual countries. Further, as we are interested in vulnerabilities
prior to crises and do not lag explanatory variables for this purpose,
the benchmark dependent variable is defined as a specified number
of years prior to the crisis. In the horse race, the benchmark is
5--12 quarters prior to a crisis.

As proposed by \citet{Bussiere2006953}, we account for post-crisis
and crisis bias by not including periods when a crisis occurs or the
two years thereafter. The excluded observations are not informative
regarding the transition from tranquil times to distress events, as
they can neither be considered \textquotedblleft normal\textquotedblright{}
periods nor vulnerabilities prior to crises. Following the same reasoning,
observations 1--4 quarters prior to crises are also left out. To issue
binary signals with method $m$, we need to specify a threshold value
$\tau$ on the estimated probabilities $p_{n}^{m}$, which is set
as to optimize Usefulness (as outlined in Section \ref{sub:Early-warning-as}).
We assume a policymaker to be more concerned of missing a crisis than
giving a false alarm. Hence, the benchmark preference $\mu$ is assumed
to be 0.8. This reasoning follows the fact that a signal is treated
as a call for internal investigation, whereas significant negative
repercussions of a false alarm only descend from external announcements.

For comparability, we consistently transform output probabilities
of each method into their own percentile distributions of in-sample
data. This is particularly relevant for model aggregation, as it is
important for model output to be on the same scale. More specifically,
the empirical cumulative distribution function is computed based on
the in-sample probabilities for each method, and both the in-sample
and out-of-sample probabilities are converted to percentiles of the
in-sample probabilities.

\paragraph{Estimation strategies}

With the aim of tackling the classification problem at hand, this
paper uses two conceptually different estimation strategies. First,
we use cross-validation for preventing overfitting and for objective
comparisons of generalization performance. Second, we test the performance
of methods when applied in the manner of a real-time exercise.

The resampling method of cross-validation, as introduced by \citet{Stone1977}
in the 1970s, is commonly used in machine learning to assess the generalization
performance of a model on out-of-sample data and to prevent overfitting.
Out of a range of different approaches to cross-validation, we make
use of so-called $K$-fold cross-validation. In line with the famous
evidence by \citet{Shao1993}, leave-\LyXZeroWidthSpace \LyXZeroWidthSpace one-\LyXZeroWidthSpace \LyXZeroWidthSpace out
cross-validation does not lead to a consistent estimate of the underlying
true model, whereas certain kinds of leave-\LyXZeroWidthSpace \LyXZeroWidthSpace \emph{n}-\LyXZeroWidthSpace \LyXZeroWidthSpace out
cross-\LyXZeroWidthSpace \LyXZeroWidthSpace validation are consistent.
Further, \citet{Breiman1996b} shows that leave-one-out cross-validation
may also run into trouble with the problem that a small change in
the data causes a large change in the model selected, whereas \citet{BreimanSpector1992}
and \citet{Kohavi1995} found that $K$-fold works better than leave-one-out
cross-validation. For an extensive survey article on cross-validation
see \citet{ArlotCelisse2010}. Cross-validation is used here in two
ways. The first aim of cross-validation is to function as a tool for
model selection for obtaining optimal free parameters, with the aim
of generalizing data rather than (over)fitting on in-sample data.
The other aim relates to objective comparisons of models' performance
on out-of-sample data, given an identical sampling for the cross-validated
model estimations. The scheme used herein involves sampling data into
$K$ folds for cross-validation and functions as follows:
\begin{enumerate}
\item Randomly split the set of observations into $K$ folds of approximately
equal size.
\item For the $k\mbox{th}$ out-of-sample validation fold, fit a model to
and compute an optimal threshold $\tau^{*}$ using $U_{r}^{K-1}(\mu)$
with the remaining $K-1$ folds, also called the in-sample data. Apply
the threshold to the $k\mbox{th}$ fold and collect its out-of-sample
$U_{r}^{k}(\mu)$.
\item Repeat Steps 1 and 2 for $k=1,2,...,K$, and collect out-of-sample
performance for all $K$ validation sets as $U_{r}^{K}(\mu)=\frac{1}{K}\sum_{k=1}^{K}U_{r}^{k}(\mu)$.\footnote{This is only a simplification of the precise implementation. We in
fact sum up all elements of the contingency matrix, and only then
compute a final Usefulness $U_{r}^{K}(\mu)$. }
\end{enumerate}
For model selection, a grid search of free parameters is performed
for the methods supporting those. As stated previously, $K$-fold
cross-validation is used and the free parameters yielding the best
performance on the out-of-sample data are stored and applied in subsequent
analysis. The literature has generally preferred small values for
$K$, with $K=10$ being among the most prominently used number of
folds (see e.g. \citet{Zhang1993}.) The cross-validated horse race
makes use of 10-fold cross-validation to provide objective relative
assessments of generalization performance of different models. The
latter purpose of cross-validation is central for the horse race,
as it allows for comparisons of models, and thus different modeling
techniques, but still assures identical sampling.

The standard approach to cross-validation may not, however, be entirely
unproblematic. As we make use of panel data, including a cross-sectional
and time dimension, we should also account for the fact that the data
are more likely to exhibit temporal dependencies. Although the cross-validation
literature has put forward advanced techniques to decrease the impact
of dependence, such as a so-called modified cross-validation by \citet{ChuMarron1991}
(further examples in \citet{ArlotCelisse2010}), the most prominent
approach is to limit estimation samples to historical data for each
prediction. To test models from the viewpoint of real-time analysis,
we use a recursive exercise that derives a new model at each quarter
using only information available up to that point in time.\footnote{It is worth noting that it is still well-motivated to use two separate
tests, cross-validated and recursive evaluations. If we would also
optimize free parameters with respect to recursive evaluations, then
we might risk overfitting them to the specific case at hand. Thus,
in case optimal parameters chosen with cross-validation also perform
in recursive evaluations, we can assure that models are not overfitting
data.} This enables testing whether the use of a method would have provided
means for predicting the global financial crisis of 2007--2008, and
how methods are ranked in terms of performance for the task. This
involves accounting for publication lags by lagging accounting based
measures with 2 quarters and market-based variables with 1 quarter.
The recursive algorithm proceeds as follows. We estimate a model at
each quarter with all available information up to that point, evaluate
the signals to set an optimal threshold $\tau^{*}$, and provide an
estimate of the current vulnerability of each economy with the same
threshold as on in-sample data. The threshold is thus time-varying.
At the end, we collect all probabilities and thresholds, as well as
the signals, and evaluate how well the model has performed in out-of-sample
analysis. As any ex post assessment, it is crucial to acknowledge
that also this exercise is performed in a quasi real-time manner with
the following caveats. Given how data providers report data, it is
not possible to account for data revisions, and potential changes
may hence have occurred after the first release. Moreover, we experiment
with two different approaches for real-time use of pre-crisis periods
as the dependent variable. With a forecast horizon of three years,
we will at each quarter know with certainty only after three years
whether or not the current quarter is a pre-crisis period to a crisis
event (unless a crisis has occurred in the past three years). We test
both dropping a window of equal length as the forecast horizon and
using pre-crisis periods for the assigned quarters.\footnote{Drawbacks of dropping a pre-crisis window are that it would require
a much later starting date of the recursion due to the short time
series and that it would distort the real relationship between indicators
and pre-crisis events. The latter argument implies that model selection,
particularly variable selection, with dropped quarters would be biased.
For instance, if one indicator perfectly signals all simultaneous
crises in 2008, but not earlier crises, a recursive test would show
bad performance, and point to concluding that the indicator is not
useful. In contrast to lags on the independent variables, which impact
the relationship to the dependent variable, it is worth noting that
using the approach with pre-crisis periods does not impact the latest
available relationship in data and information set at each quarter.} As a horse race, the recursive estimations test the models from the
viewpoint of real-time analysis. Using in-sample data ranging back
as far as possible, the recursive exercise starts from 2005Q2, with
the exception of the QDA method, for which analysis starts from 2006Q2,
due to requirements of more training data than for the other methods.
This procedure enables us to test performance with no prior information
on the build-up phase of the recent crisis.

\subsection{Aggregation procedures\label{sub:Aggregation-procedures}}

From individual methods, we move forward to combining the outputs
of several different methods into one through a number of aggregation
procedures. The approaches here descend from the subfield of machine
learning focusing on ensemble learning, wherein the main objective
is the use of multiple statistical learning algorithms for better
predictive performance. Although we aim for simplicity and do not
adopt the most complex algorithms herein, we make use of the two common
approaches in ensemble learning: bagging and boosting. \emph{Bagging}
stands for Bootstrap Aggregation \citep{Breiman1996} and makes use
of resampling from the original data, which is to be aggregated into
one model output. While being an approach for ensemble learning,
we discuss this under the topic of resampling and model uncertainty,
as can be seen in Section \ref{sub:Model-uncertainty}. \emph{Boosting
}\citep{Schapire1990} refers to computing output from multiple models
and then averaging the results with specified weights, which we mainly
rely on in our aggregation procedures below. A third group of stacking
approaches \citep{Wolpert1992}, which add another layer of models
on top of individual model output to improve performance, are not
used in this paper for the sake of simplicity. Again, we use the optimal
free parameters identified through cross-validated grid searches,
and then estimate individual methods. For this, we make use of four
different aggregation procedures: the best-of and voting approaches,
and arithmetic and weighted averages of probabilities. 

The best-of approach simply makes use of one single method $m$ by
choosing the most accurate one. To use information in a truthful manner,
we always choose the method, independent of the exercise (i.e., cross-validation
or recursion), which has the best in-sample relative Usefulness. Voting
simply makes use of the signals $B_{n}^{m}$ of all methods $m=1,2,...,M$
for each observation $x_{n}$ in order to signal or not based upon
a majority vote. That is, the aggregate $B_{n}^{a}$ chooses for observation
$x_{n}$ the class that receives the largest total vote from all individual
methods:

\[
B_{n}^{a}=\left\{ \begin{array}{ccc}
1 & \mbox{if} & \frac{1}{M}\sum_{m=1}^{M}B_{n}^{m}>0.5\\
0 & \mbox{otherwise}
\end{array}\right.,
\]
where $B_{n}^{m}$ is the binary output for method $m$ and observation
$n$, and $B_{n}^{a}$ is the binary output of the majority vote aggregate.

Aggregating probabilities requires an earlier intervention in modeling.
In contrast to the best-of and voting approaches, we directly make
use of the probabilities $p_{n}^{m}$ of each method $m$ for all
observations $n$ to average them into aggregate probabilities. The
simpler case uses an arithmetic mean to derive aggregate probabilities
$p_{n}^{a}$. For weighted aggregate probabilities $p_{n}^{a}$, we
make use of in-sample model performance when setting the weights of
methods, so that the most accurate method (in-sample) is given the
most weight in the aggregate. The non-weighted and weighted probabilities
$p_{n}^{a}$ for observations $x_{n}$ can be derived as follows:

\[
p_{n}^{a}=\sum_{m=1}^{M}\frac{w_{m}}{\sum_{m=1}^{M}w_{m}}p_{j}^{m}
\]
where the probabilities $p_{n}^{m}$ of each method $m$ are weighted
with its performance measure $w_{m}$ for all observations $n$. In
this paper, we make use of weights $w_{m}=U_{r}^{m}(\mu)$, but the
approach is flexible for any chosen measure, such as the AUC. This
weighting has the property of giving the least useful method the smallest
weight, and accordingly a bias towards the more useful ones. The arithmetic
mean can be shown to result in $p_{n}^{a}=\frac{1}{M}\sum_{m=1}^{M}p_{n}^{m}$
for $w_{m}=1$. To make use of only available information in a real-time
set-up, the $U_{r}^{m}(\mu)$ used for weighting refers always to
in-sample results. In order to assure non-negative weights, we drop
methods with negative values (i.e., $U_{r}^{m}(\mu)<0$) from the
vector of performance measures. In the event that all methods show
a negative Usefulness, they are given weights of equal size. After
computing aggregate probabilities $p_{n}^{a}$, they are treated as
if they were outputs for a single method (i.e., $p_{n}^{m}$), and
optimal thresholds $\tau^{*}$ identified accordingly. In contrast,
the best-of approach signals based upon the identified individual
method and voting signals if and only if a majority of the methods
signal, which imposes no requirement of a separate threshold. Thus,
the overall cross-validated Usefulness of the aggregate is calculated
in the same manner as for individual methods. Likewise, for the recursive
model, the procedure is identical, including the use of in-sample
Usefulness $U_{r}^{m}(\mu)$ for weighting.

\subsection{Model uncertainty\label{sub:Model-uncertainty}}

We herein tackle uncertainty in classification tasks concerning model
performance uncertainty and model output uncertainty. While descending
from multiple sources and relating to multiple features, we are particularly
concerned with uncertainties coupled with model parameters.\footnote{Beyond model parameter uncertainty, and no matter how precise the
estimates are, models will not be perfect and hence there is always
a residual model error. To this end, we are not tackling uncertainty
in model output (or model error) resulting from errors in the model
structure, which particularly relates to the used crisis events and
indicators in our dataset (i.e., independent and dependent variables).} Accordingly, we assess the extent to which model parameters, and
hence predictions, vary if models were estimated with different datasets.
With varying data variation in the predictions is caused by imprecise
parameter values, as otherwise predictions would always be the same.
Not to confuse variability with measures of model performance, zero
parameter value uncertainty in the predictions would still not imply
perfectly accurate predictions. To represent any uncertainty, we need
to derive properties of the estimates, including standard errors (SEs),
confidence intervals (CIs) and critical values (CVs). To move toward
robust statistical analysis in early-warning modeling, we first present
our general approach to early-warning inference through resampling,
and then present the required specification for assessing model performance
and output uncertainty.

\paragraph{Early-warning inference}

The standard approaches to inference and deriving properties of estimates
descend from conventional statistical theory. If we know the data
generating process (DGP), we also know that for data $x_{1},x_{2},...,x_{N}$,
we have the mean $\hat{\theta}=\sum_{n=1}^{N}x_{n}/N$ as an estimate
of the expected value of $x$, the SE $\hat{\sigma}=\sqrt{\sum_{n=1}^{N}\left(x_{n}-\hat{\theta}\right)^{2}/N^{2}}$
showing how well $\hat{\theta}$ estimates the true expectation, and
the CI through $\hat{\theta}\pm t\cdot\hat{\sigma}$ (where $t$ is
the CV). Yet, we seldom do know the DGP, and hence cannot generate
new samples from the original population. In the vein of the above
described cross-validation \citep{Stone1977}, we can generally mimic
the process of obtaining new data through the family of resampling
techniques, including also permutation tests \citep{Fisher1935},
the jackknife \citep{Quenouille1949} and bootstraps \citep{Efron1979}.
At this stage, we broadly define resampling as random and repeated
sampling of sub-samples from the same, known sample. Thus, without
generating additional samples, we can use the sampling distribution
of estimators to derive the variability of the estimator of interest
and its properties (i.e., SEs, CIs and CVs). For a general discussion
of resampling techniques for deriving properties of an estimator,
the reader is referred to original works by \citet{Efron1981,Efron1992}
and \citet{EfronTibshirani1986,EfronTibshirani1993}.

Let us consider a sample with $n=1,...,N$ independent observations
of one dependent variable $y_{n}$ and $G+1$ explanatory variables
$x_{n}$. We consider our resamplings to be paired by drawing independently
$N$ pairs $(x_{n},y_{n})$ from the observed sample. Resampling involves
drawing randomly samples $s=1,...,S$ from the observed sample, in
which case an individual sample is ($x_{n}^{s},y_{n}^{s}$). To estimate
SEs for any estimator $\hat{\theta}$, we make use of the empirical
standard deviation of resamplings $\hat{\theta}$ for approximating
the SE $\sigma(\hat{\theta})$. We proceed as follows:
\begin{enumerate}
\item Draw $S$ independent samples ($x_{n}^{s},y_{n}^{s}$) of size $N$
from $(x_{n},y_{n})$.
\item Estimate the parameter $\theta$ through $\hat{\theta}_{s}^{*}$ for
each resampling $s=1,...,S$. 
\item Estimate $\sigma(\hat{\theta})$ by $\hat{\sigma}=\sqrt{\frac{1}{S-1}\sum_{s=1}^{S}\left(\hat{\theta}_{s}^{*}-\hat{\theta}^{*}\right)^{2}}$,
where $\hat{\theta}^{*}=\frac{1}{S}\sum_{s=1}^{S}\hat{\theta}_{s}^{*}$.
\end{enumerate}
Now, given a consistent and asymptotically normally distributed estimator
$\hat{\theta}$, the resampled SEs can be used to construct approximate
CIs and to perform asymptotic tests based on the normal distribution,
respectively. Thus, we can use percentiles to construct a two-sided
asymmetric but equal-tailed $(1-\alpha)$ CI, where the empirical
percentiles of the resamplings ($\alpha/2$ and $1-\alpha/2$) are
used as lower and upper limits for the confidence bounds. We make
use of the above Steps 1 and 2, and then proceed instead as follows:
\begin{enumerate}
\item [4.]Order the resampled replications of estimator $\hat{\theta}$
such that $\hat{\theta}_{1}^{*}\leq...\leq\hat{\theta}_{B}^{*}$.
With the $S\cdot\alpha/2$th and $S\cdot(1-\alpha/2)$th ordered elements
as the lower and upper limits of the confidence bounds, the estimated
$(1-\alpha)$ CI of $\hat{\theta}$ is $\left[\hat{\theta}_{S\cdot\alpha/2}^{*},\hat{\theta}_{S\cdot(1-\alpha/2)}^{*}\right]$.
\end{enumerate}
Using the above discussed resampled SEs and approximate CI, we can
use a conventional (but approximate) two-sided hypothesis test of
the null $H_{0}:\theta=\theta^{0}$. In case $\theta^{0}$ is outside
the two-tailed $(1-\alpha)$ CI with the significance level $\alpha$,
the null hypothesis is rejected. Yet, if we have two resampled estimators
$\hat{\theta}^{i}$ and $\hat{\theta}^{j}$ with non-overlapping CIs,
it is obvious that they are necessarily significantly different, but
it is not necessarily true that they are not significantly different
if they overlap. Rather than mean CIs, we are concerned with the test
statistic for the difference between two means. Two means are significantly
different for $(1-\alpha)$ confidence levels when the CI for the
difference between the group means does not contain zero: $\left(\hat{\theta}^{i}-\hat{\theta}^{j}\right)-t\sqrt{\hat{\sigma}_{i}^{2}+\hat{\sigma}_{j}^{2}}>0$.\footnote{In contrast to the test statistic, we can see that two means have
no overlap if the lower bound of the CI for the greater mean is greater
than the upper bound of the CI for the smaller mean, or $\hat{\theta}^{i}+t\cdot\hat{\sigma}^{i}>\hat{\theta}^{j}+t\cdot\hat{\sigma}^{j}$.
While simple algebra gives that there is no overlap if $\hat{\theta}^{i}-\hat{\theta}^{j}>t\left(\hat{\sigma}^{i}+\hat{\sigma}^{j}\right)$,
the test statistic only differs through the square root and the sum
of squares: $\hat{\theta}^{i}-\hat{\theta}^{j}>t\sqrt{\hat{\sigma}_{i}^{2}+\hat{\sigma}_{j}^{2}}$.
As $\sqrt{\hat{\sigma}_{i}^{2}+\hat{\sigma}_{j}^{2}}<\hat{\sigma}^{i}+\hat{\sigma}^{j}$,
it is obvious that the mean difference becomes significant before
there is no overlap between the two group-mean CIs.} Yet, we may be violating the normality assumption as the traditional
Student $t$ distribution for calculating CIs relies on a sampling
from a normal population.

Even though we could by the central limit theorem argue for the distributions
to be approximately normal if the sampling of the parent population
is independent, the degree of the approximation would still depend
on the sample size $N$ and on how close the parent population is
to the normal. As the common purpose behind resampling is not to impose
such distributional assumptions, a common approach is to rely on so-called
resampled $t$ intervals. Thus, based upon statistics of the resamplings,
we can solve for $t^{*}$ and use confidence cut-offs on the empirical
distribution. Given consistent estimates of $\hat{\theta}$ and $\hat{\sigma}(\hat{\theta})$,
and a normal asymptotic distribution of the $t$-statistic $t=\frac{\hat{\theta}-\theta_{0}}{\hat{\sigma}(\hat{\theta})}\rightarrow N(0,1)$,
we can derive approximate symmetrical CVs $t^{*}$ from percentiles
of the empirical distribution of all resamplings for the $t$-statistic.
\begin{enumerate}
\item Consistently estimate the parameter $\theta$ and $\sigma(\hat{\theta})$
using the observed sample: $\hat{\theta}$ and $\hat{\sigma}(\hat{\theta})$.
\item Draw $S$ independent resamplings ($x_{n}^{s},y_{n}^{s}$) of size
$N$ from $(x_{n},y_{n})$.
\item Assuming $\theta^{0}=\hat{\theta}$, estimate the $t$-value $t_{s}^{*}=\frac{\hat{\theta}_{s}^{*}-\hat{\theta}}{\hat{\sigma}_{s}^{*}(\hat{\theta})}$
for $s=1,...,S$ where $\hat{\theta}_{s}^{*}$ and $\hat{\sigma}_{s}^{*}(\hat{\theta})$
are resampled estimates of $\theta$ and its SE.
\item Order the resampled replications of $t$ such that $\left|t_{1}^{\text{\textasteriskcentered}}\right|\le...\le\left|t_{S}^{\text{\textasteriskcentered}}\right|$.
With the $S\cdot\left(1-\alpha\right)$th ordered element as the CV,
we have $t_{\alpha/2}=\left|t_{S\cdot\left(1-\alpha\right)}^{*}\right|$
and $t_{1-\alpha/2}=\left|t_{S\cdot\left(1-\alpha\right)}^{*}\right|$.
\end{enumerate}
With these symmetrical CVs, we can utilize the above described mean-comparison
test. Yet, as CVs for the resampled $t$ intervals may differ for
the two means, we amend the test statistic as follows:

\[
\left(\hat{\theta}^{i}-\hat{\theta}^{j}\right)-\frac{t_{S\cdot\left(1-\alpha\right)}^{*j}+t_{S\cdot\left(1-\alpha\right)}^{*i}}{2}\sqrt{\hat{\sigma}_{i}^{2}+\hat{\sigma}_{j}^{2}}>0.
\]

\paragraph{Model performance uncertainty}

For a robust horse race, and ranking of methods, we make use of resampling
techniques to assess variability of model performance. We compute
for each individual method and the aggregates resampled SEs for the
relative Usefulness and AUC measures. Then, we use the SEs to obtain
CVs for the measures, analyze pairwise among methods and aggregates
whether intervals exhibit statistically significant overlaps, and
produce a matrix that represents pairwise significant differences
among methods and aggregates. More formally, the null hypothesis that
methods $i$ and $j$ have equal out-of-sample performance can be
expressed as $H_{0}:\,U_{r}^{i}(\mu)=U_{r}^{j}(\mu)$ (and likewise
for AUC). To this end, the alternative hypothesis of a difference
in out-of-sample performance of methods $i$ and $j$ is $H_{1}:\,U_{r}^{i}(\mu)\neq U_{r}^{j}(\mu)$.

In machine learning, supervised learning algorithms are said to be
prevented from generalizing beyond their training data due to two
sources of error: bias and variance. While bias refers to error from
erroneous assumptions in the learning algorithm (i.e., underfit),
variance relates to error from sensitivity to small fluctuations in
the training set (i.e., overfit). The above described $K$-fold cross-validation
may run the risk of leading to models with high variance and non-zero
yet small bias (e.g., \citet{Kohavi1995,Hastieetal2011}). To address
the possibility of a relatively high variance and to better derive
estimates of properties (i.e., SEs, CIs and CVs), repeated cross-validations
are oftentimes advocated. This allows averaging model performance,
and hence ranking average performance rather than individual estimations,
as well as better enables deriving properties of the estimates.\footnote{Repeated cross-validations are not entirely unproblematic (e.g., \citet{VanwinckelenBlockeel2012}),
yet still one of the better approaches to simultaneously assess generalizability
and uncertainty.} For both individual methods and aggregates, we make use of 500 repetitions
of the cross-validations (i.e., $S=500$).

In the recursive exercises, we opt to make use of resampling with
replacement to assess model performance uncertainty due to limited
sample sizes for the early quarters. The family of bootstrapping approaches
was introduced by \citet{Efron1979} and \citet{EfronTibshirani1993}.
Given data $x_{1},x_{2},...,x_{N}$, bootstrapping implies drawing
a random sample of size $N$ through resampling with replacement from
$x$, leaving some data points out while others will be duplicated.
Accordingly, an average of roughly 63\% of the training data is utilized
for each bootstrap. However, the standard bootstrap procedure assumes
data to be i.i.d., and thus does not account for possible dependencies
present in the data. Since early-warning models commonly use panel
data, both cross-sectional and time-series dependence are to be assumed.
In line with \citet{Kapetanios2008} and \citet{Hounkannounon2011},
we thus utilize a double bootstrap for the robust recursive horse
race, consisting of two components: cross-sectional resampling and
the moving block bootstrap. For panel data of dimensions $E\times T$,
where $E$ is the number of entities, and $T$ is the number of periods,
cross-sectional resampling entails drawing full time-series for $E$
entities with replacement. The moving block bootstrap, introduced
by \citet{Kunsch1989}, draws blocks of a defined size $B$ of observations,
in order to preserve temporal dependency within the resampled blocks.
Our double bootstrap procedure combines both in the following way: 
\begin{enumerate}
\item From the available in-sample data of dimensions $E\times N$ , draw
$E$ entities with replacement. This constitutes the pseudo-sample
$S^{*}$. 
\item From the obtained pseudo-sample $S^{*}$, draw a randomly selected
block of size $B$ from all $E$ entities. 
\item Repeat 2. until the length of all combined blocks is > $N$ by cutting
at the end. This constitutes the final bootstrap sample $S^{**}.$
\end{enumerate}
For each quarter, we draw randomly the bootstrap sample $S^{**}$
from the available in-sample data using the above procedure, which
is repeated 500 times. Each of these bootstraps are treated individually
to compute the performance of individual methods and the aggregates.
These results are then averaged to obtain the corresponding results
of a robust bootstrapped classifier for each method and aggregate.

\paragraph{Model output uncertainty}

In order to assess the reliability of estimated probabilities and
optimal thresholds, and hence signals, we study the notion of model
output uncertainty. The question of interest would be whether or not
an estimated probability is statistically significantly above or below
a given optimal threshold. More formally, the null hypothesis that
probabilities $p_{n}\in\left[0,1\right]$ and optimal thresholds $\tau_{n}^{*}\in\left[0,1\right]$
are equal can be expressed as $H_{0}:\,p_{n}=\tau_{n}^{*}$. Hence,
the alternative hypothesis of a difference in probabilities $p_{n}$
and optimal thresholds $\tau_{n}^{*}$ is $H_{1}:\,p_{n}\neq\tau_{n}^{*}$.
This can be tested both for probabilities of individual methods $p_{n}^{m}$
and probabilities of aggregates $p_{n}^{a}$ as well as their thresholds
$\tau_{n}^{*m}$ and $\tau_{n}^{*a}$.

We assess the trustworthiness of the output of models, be they individual
methods or aggregates, by computing SEs for the estimated probabilities
and the optimal thresholds. We follow the approach for model performance
uncertainty to compute CVs and mean-comparison tests. For both cross-validation
and bootstraps, the 500 resamplings of the out-of-sample probabilities
are computed separately for each method and averaged with and without
weighting, as above discussed (i.e., $S=500$). From these, the mean
and the SE are drawn and used to construct a CV for individual methods
and the aggregates, based on bootstrapped crisis probabilities and
optimal thresholds, which allows us to test when model output is statistically
significantly above or below a threshold. The above implemented bootstraps
also serve another purpose. We make use of the CI as a visual representation
of uncertainty. Thus, we produce confidence bands $\left[\hat{\theta}_{S\cdot\alpha/2}^{*},\hat{\theta}_{S\cdot(1-\alpha/2)}^{*}\right]$
around time-series of probabilities and thresholds for each method
and country, which is useful information for policy purposes when
assessing the reliability of model output.

\subsection{Summary of horse race exercises}

To sum up the above described exercises, we herein provide a simplified
description of the cross-validated and the recursive horse races,
as well as steps within them.
\begin{itemize}
\item Cross-validation: Split the full sample into $k$ folds of equal size,
and estimate models and thresholds using the remaining $k-1$ folds
of data. Collect out-of-sample probabilities and binary predictions
for each left-out fold.
\item Recursive: Utilize an out-of-sample span split into individual quarters,
for which the model is estimated and optimal threshold computed using
all data available up until each quarter. 
\end{itemize}
For both exercises, all out-of-sample output is finally reassembled
and performance summarized in terms of a range of evaluation measures.
The two exercises differ in their sampling of data, particularly the
in-sample and out-of-sample partitions used for each estimation. While
cross-validation is common in machine learning and allows an efficient
use of small samples, exercises may benefit from the fact that data
are sampled randomly despite most likely exhibiting time dependence.
The recursive exercises, on the contrary, account for time dependence
in data by strictly using historical samples for out-of-sample predictions,
which nevertheless requires more data, particularly in the time-series
dimension. These two exercises allow exploring performance across
methods, and how that is impacted by the evaluation exercise.

For both exercises, we go through the following steps to estimate
individual models, aggregate model output and represent model and
performance uncertainty:
\begin{itemize}
\item Following the above exercises, estimate models with all individual
methods $m=1,2,...,M$.
\item Aggregate model output $p^{m}$ from $M$ models to $p^{a}$ using
four approaches: best-of, voting, non-weighted and weighted.
\item Represent model performance uncertainty for individual and aggregated
methods by repeating the exercises using sampling of in-sample data
with and without replacement and reporting statistically significant
rankings.
\item Represent model output uncertainty for individual and aggregated methods
by repeating the exercises using sampling of in-sample data with and
without replacement and reporting statistically significant signals
and non-signals.
\end{itemize}

\section{The European crisis as a playground}

This section applies the above introduced concepts in practice. Using
a European sample, we implement the horse race with a large number
methods, apply aggregation procedures and illustrate the use and usefulness
of accounting for and representing model uncertainty.

\subsection{Model selection}

To start with, we need to derive suitable (i.e., optimal) values for
the free parameters for a number of methods. Roughly half of the above
discussed methods have one or more free parameters relating to their
learning algorithm, for which optimal values are identified empirically.
In summary, these methods are: signal extraction, LASSO, KNN, classification
trees, random forest, ANN, ELM and SVM. To perform model selection
for these six methods, we make use of a grid search to find optimal
free parameters with respect to out-of-sample performance. A set of
tested values are selected based upon common rules of thumb for each
free parameter (i.e., usually minimum and maximum values and regular
steps in between), whereafter an exhaustive grid search is performed
on the discrete parameter space of the Cartesian product of the parameter
sets. To obtain generalizable models, we use 10-fold cross-validation
and optimize out-of-sample Usefulness for guiding the specifications
of the algorithms. Finally, the parameter combinations yielding the
highest out-of-sample Usefulness are chosen, as is optimal for each
method. For the signal extraction method, we vary the used indicator,
and the indicator with the highest Usefulness is chosen (for a full
table see Table \ref{tab:Signalextraction} in the Appendix).\footnote{As the poor performance of signal extraction may arise questions,
we also show results for $\mu=0.9193=1-\mbox{Pr\ensuremath{(C=1)}}$
in Table \ref{tab:Signalextraction-mu0.91} in the Appendix. Given
the unconditional probabilities of events, this preference parameter
has potential to yield the largest Usefulness. Accordingly, we can
also find much larger Usefulness values for most indicators. This
highlights the sensitivity of signal extraction to the chosen preferences.} The chosen parameters are reported in Table \ref{tab:Optimal-parameters-obtained}.\footnote{It may be noted the the optimal amount of hidden units for the ELM
method returned by the grid-search algorithm is unusually high. However,
as seen below in the cross-validated and particularly real-time exercises,
the results obtained using the ELM method do not seem to exhibit overfitting.
Also, by comparing results of the ELM to those of the ANN, which has
only eight hidden units, out-of-sample results are in all tests similar
in nature.}

\begin{table}[H]
\centering{}\protect\caption{\label{tab:Optimal-parameters-obtained}Optimal parameters obtained
through a grid-search algorithm.}
{\footnotesize{}}%
\begin{tabular}{cccc}
{\footnotesize{}Method} & \multicolumn{3}{c}{{\footnotesize{}Parameters}}\tabularnewline
\hline 
\hline 
{\footnotesize{}Signal extraction} & {\footnotesize{}Debt service ratio} &  & \tabularnewline
{\footnotesize{}Logit LASSO} & {\footnotesize{}$\lambda=0.0012$} &  & \tabularnewline
{\footnotesize{}KNN} & {\footnotesize{}$k$ = 2} & {\footnotesize{}Distance = 1} & \tabularnewline
{\footnotesize{}Trees} & {\footnotesize{}Complexity = 0.01} &  & \tabularnewline
{\footnotesize{}Random forest} & {\footnotesize{}No. of trees = 180} & {\footnotesize{}No. of predictors sampled = 5} & \tabularnewline
{\footnotesize{}ANN} & {\footnotesize{}No. of hidden layer units = 8} & {\footnotesize{}Max no. of iterations = 200} & {\footnotesize{}Weight decay = 0.005}\tabularnewline
{\footnotesize{}ELM} & {\footnotesize{}No. of hidden layer units = 300} & {\footnotesize{}Activation function = Tan-sig} & \tabularnewline
{\footnotesize{}SVM} & {\footnotesize{}$\gamma$ = 0.4} & {\footnotesize{}Cost = 1} & {\footnotesize{}Kernel = Radial basis}\tabularnewline
\end{tabular}{\footnotesize \par}
\end{table}

\subsection{A horse race of early-warning models}

We conduct in this section two types of horse races: a cross-validated
and a recursive. This provides a starting point for the ranking of
early-warning methods and simultaneous use of multiple models.

\paragraph{Cross-validated race}

The first approach to ranking early-warning methods uses 10-fold cross-validation.
Rather than optimizing free parameters, the cross-validation exercise
aims at producing comparable models with all included methods, which
can be assured due to the similar sampling of data and modeling specifications.
For the above discussed methods, we use the optimal parameters as
shown in Table \ref{tab:Optimal-parameters-obtained}. Methods with
no free parameters are run through the 10-fold cross-validation without
any further ado. Table \ref{tab:A-horse-race-} presents the out-of-sample
results of the cross-validation horse race for the individual early-warning
methods, sorted by descending Usefulness. At first, we can note that
the simple approaches, such as signal extraction, LDA and logit analysis,
are outperformed in terms of Usefulness by most machine learning techniques.
At the other end, the methods with highest Usefulness are KNN and
SVM. In terms of AUC, QDA, random forest, ANN, ELM and SVM yield good
results. It is still worth noting that a standard cross-validated
test does not account for potential excessive correlation across folds
due to dependence in data, and hence the more flexible non-linear
approaches are also more prone to exhibit a too good model fit. Yet,
this can easily be controlled for with the recursive real-time analysis.

\begin{table}[H]
\protect\caption{\label{tab:A-horse-race-}A horse race of cross-validated estimations.}

\centering{}\includegraphics[width=1\textwidth]{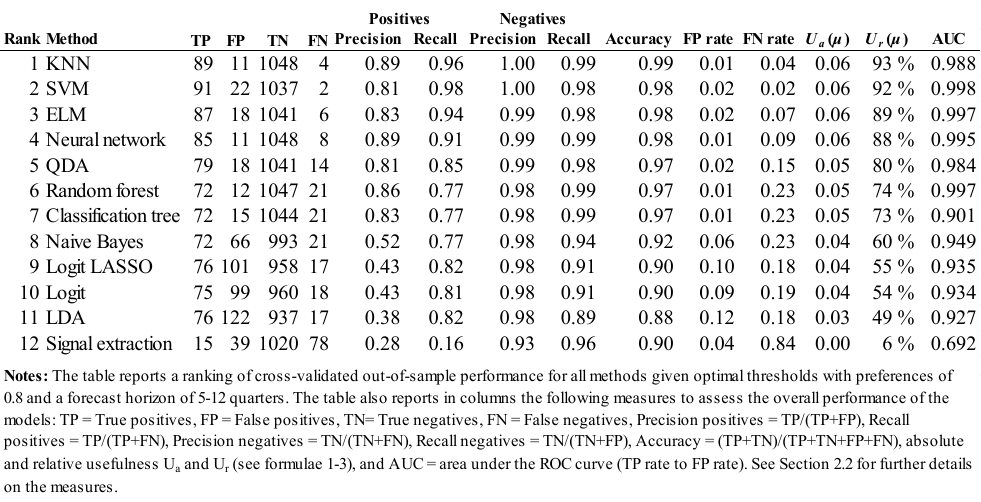}
\end{table}

\paragraph{Recursive race}

To further test the performance of all individual methods, we conduct
a recursive horse race among the approaches. As outlined in Section
\ref{sub:Set-up-of-the-horse-race}, we estimate new models with the
available information in each quarter to identify vulnerabilities
in the same quarter, starting from 2005Q2 (2006Q2 for QDA). Besides
for a few exceptions, the results in Table \ref{tab:Recursive comparison}
are in line with those in the cross-validated horse race in Table
\ref{tab:A-horse-race-}. For instance, the top six methods are the
same with only minor differences in ranks, and classification trees
perform poorly in the recursive exercise and logit in the cross-validated
exercise. Generally, machine learning based approaches again outperform
more conventional techniques from the early-warning literature.

We also experiment with so-called ``unknown events'' in recursive
exercises, as any given quarter is known to be tranquil only when
the forecast horizon has passed. Hence, we test two approaches: (\emph{i})
dropping a window of equal length as the forecast horizon at each
quarter, and (\emph{ii}) simply using pre-crisis periods for the assigned
quarters. We can conclude that dropping quarters had no impact on
the ranking of methods and only minor negative impact on the levels
of performance measures. Besides for a starting quarter only in 2005Q3
due to data requirements (and only 2006Q2 for QDA), Table \ref{tab:Recursive comparison-ROB}
in the Appendix shows results for a similar recursive exercise as
in Table \ref{tab:Recursive comparison}, but where a pre-crisis window
prior to each prediction quarter has been dropped. It is to be noted
that data sparsity hinders this exercise with the current set of indicators,
due to which we drop the indicator loans to income. Although the table
shows a drop in average $U_{r}$ from 46\% to 32\% and average AUC
from 0.87 to 0.86, which might also relate to dropping one indicator,
the rankings of individual methods are with a few exceptions unchanged.
The largest change in rankings occurs for QDA, but this might to a
large extent descend from the change in the starting quarter, as well
as refers only to Usefulness as AUC is close to unchanged. Moreover,
while the $U_{r}$ (AUC) drop for machine learning approaches is on
average 13 percentage points (0.01), the more conventional statistical
approaches drop by 16 percentage points (0.05). Hence, this does not
point to an overfit caused by assigning events to reference quarters.

The added value of a palette of methods is that it not only allows
for handpicking the best-in-class techniques, but also the simultaneous
use of all or a number of methods. As some of the recent machine learning
approaches may be seen as less interpretable for those unfamiliar
with them, the simultaneous use of a large number of methods may build
confidence through performance comparisons and the simultaneous assessment
of model output. The purpose of multiple models would hence relate
to confirmatory uses, as policy is most often an end product of a
discretionary process. On the other hand, the dissimilarity in model
output may also be seen as a way to illustrate uncertainty of or variation
in model output. Yet, this requires a more structured assessment (as
is done in Section \ref{sub:Model-uncertainty-1}).

\begin{table}[H]
\protect\caption{\label{tab:Recursive comparison}A horse race of recursive real-time
estimations.}

\centering{}\includegraphics[width=1\textwidth]{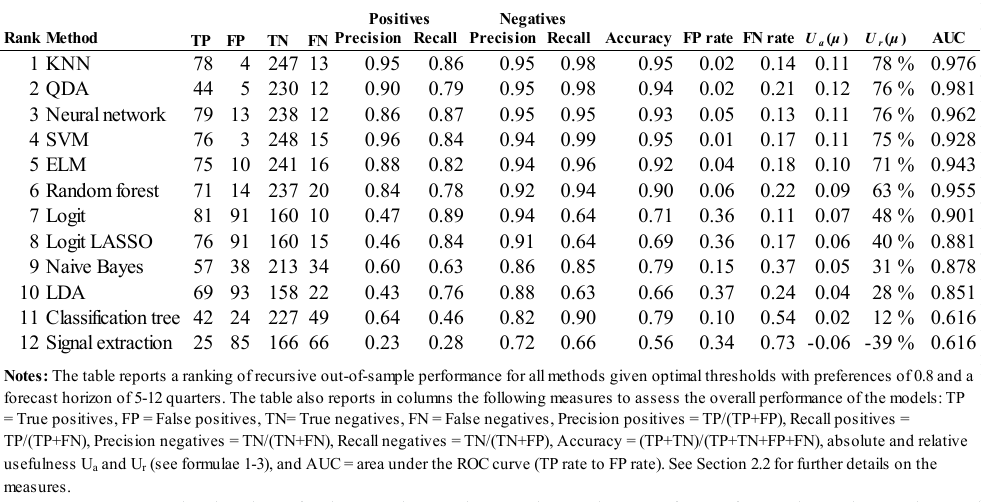}
\end{table}

\subsection{Aggregation of models}

Beyond the use of a single technique, or many techniques in concert,
the obvious next step is to aggregate them into one model output.
This is done with four approaches, as outlined in Section \ref{sub:Aggregation-procedures}.
The first two approaches combine the signals of individual methods,
by using (\emph{i}) only the best method for out-of-sample analysis
as per in-sample performance, and (\emph{ii}) a majority vote to allow
for the simultaneous use of all model signals. The third and the fourth
approach rely on the estimated probabilities for each method by deriving
an arithmetic and weighted mean of the probability for all methods
present in Tables \ref{tab:A-horse-race-} and \ref{tab:Recursive comparison}.
A natural way for weighting model output is to use their in-sample
performance, in our case relative Usefulness. This allows for giving
a larger weight to those methods that perform better and yields a
similar model output as for individual methods, which can be tested
through cross-validated and recursive exercises.

Table \ref{tab:Aggregated-results-of} presents results for four different
aggregation approaches for both the cross-validated and recursive
exercises. The simultaneous use of many models yields in general good
results. While cross-validated models rank among top five, in recursive
estimations three out of four of the aggregated approaches rank among
the best two individual approaches. One potential explanation to better
performance in the recursive exercise is that it is a more stringent
test and the cross-validated exercise might be biased through excessive
correlation among folds. Thus, when removing the potential dependence
in sampling, ensemble methods perform better than individual machine
learning methods. Further to this, we decrease uncertainty in the
chosen method, as in-sample (or a priori) performance is not an undisputed
indicator of future performance. That is, beyond the potential in
convincing policymakers' who might have a taste for one method over
others, the aggregation tackles the problem of choosing one method
based upon performance. While in-sample performance might indicate
that one method outperforms others, it might still relate to sampling
errors or an overfit to the sample at hand, and hence perform poorly
on out-of-sample data. This highlights the value of using an aggregation
rather than the choice of one single approach, however that is done.
We again experiment with so-called ``unknown events'' in recursive
exercises. Table \ref{tab:Aggregated-results-of-ROB} in the Appendix
shows similar results to those in Table \ref{tab:Aggregated-results-of}
for individual methods, when dropping unknown events in the recursive
exercise. The aggregates show a drop in average $U_{r}$ from 77\%
to 67\%, whereas the AUC on average similar. Again, no overfitting
can be observed even with the more stringent test.\footnote{Beyond having similar results, a key argument for assigning events
to the reference quarters in the sequel was that we would otherwise
need to use a later starting date of the recursion due to the short
time series.}

As can be observed in Table \ref{tab:Aggregated-results-of}, in most
cases the other aggregation approaches do not perform much better
than the results of the simple arithmetic mean. This may be related
to the fact that model diversity has been shown to improve performance
at the aggregate level (e.g., \citet{KunchevaWhitaker2003}). For
instance, more random methods (e.g., random forests) have been shown
to produce a stronger aggregate than more deliberate techniques (e.g.,
\citet{Ho1995}), in which case the aggregated models not only use
resampled observations but also resampled variables. As the better
methods of our aggregate may give similar model output, they might
lead to lesser degree of diversity in the aggregate, but it is also
worth noting that we are close to reaching perfect performance, at
which stage performance improvements obviously become more challenging.
Further approaches to ensemble learning should be a topic of future
work, as more diversity could easily be introduced to the different
learning algorithms through various approaches, such as variable and
observation resampling.

\begin{table}[H]
\protect\caption{\label{tab:Aggregated-results-of}Aggregated results of cross-validated
and recursive estimations.}

\centering{}\includegraphics[width=1\textwidth]{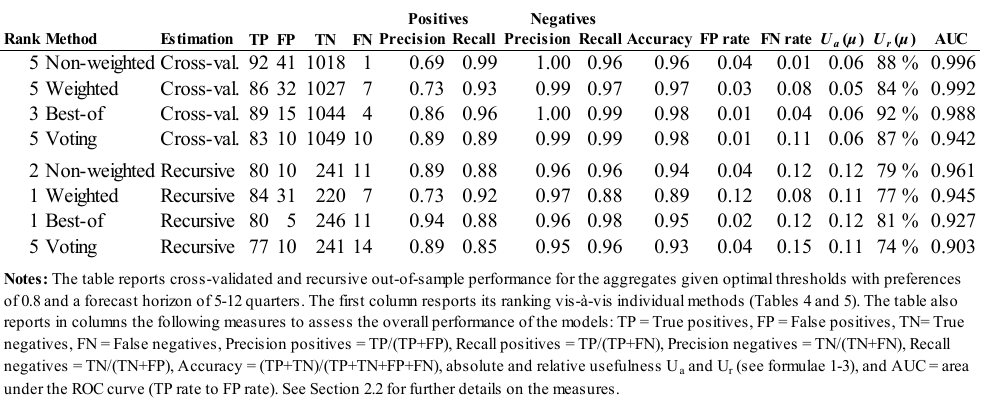}
\end{table}

\subsection{Model uncertainty\label{sub:Model-uncertainty-1}}

The final step in our empirical analysis involves computing model
uncertainty, particularly related to model performance and output.

\paragraph{Model performance uncertainty}

One may question the above horse races to be outcomes of potential
biases due to sampling error and randomness in non-deterministic methods.
This we ought to test statistically for any rank inference to be valid.
Hence, we perform similar exercises as in Tables \ref{tab:A-horse-race-},
\ref{tab:Recursive comparison} and \ref{tab:Aggregated-results-of},
but resample to account for model uncertainty. For the cross-validated
exercise, we draw 500 samples for the 10 folds, and report average
results, including SEs for three key performance measures. Thus, Table
\ref{tab:A-horse-race-ROBUST} presents a robust horse race of cross-validated
estimations. We can observe that KNN, SVM, ANN and ELM are still the
top-performing methods. They are followed by the aggregates, whereafter
the same methods as in Table \ref{tab:A-horse-race-} follow (descending
order of performance): random forests, QDA, classification trees,
logit, LASSO, LDA and signal extraction.

In addition to a simple ranking, we also use Usefulness to assess
statistical significance of rankings among all other methods. The
cross-comparison matrix for all methods can be found in Table \ref{tab:Significance_CV}
in the Appendix. The second column in Table \ref{tab:A-horse-race-ROBUST}
summarizes the results by showing the first lower-ranked method that
is statistically significantly different from each method. This indicates
clustering of model performance both among the best-in-class and worst-in-class
methods. All methods until rank 6 are shown to be better than non-weighted
aggregates ranked at number 8. Likewise, all methods above rank 11
seem to belong to a similarly performing group. The methods ranked
below the 11th have larger bilateral differences in performance, particularly
signal extraction, which is significantly poorer than all other approaches.
It is also worth noting the true ensemble approaches (i.e., aggregations
excluding the best-of approach) decrease variation in model performance,
which is expected as model averaging decreases the impact of extreme
outcomes. This is obviously of key concern when aiming at robust early-warning
models for policymaking. As a further robustness check, we also provide
cross-validated out-of-sample ROC curve plots for all methods and
the aggregates in Figure \ref{fig:AUC-CV} in the Appendix. Yet, we
prefer to focus on the Usefulness-based rankings as they focus on
a relevant point of the AUC ($\mu=0.8$), rather than covering all
potential preferences of a policymaker.

\begin{table}[H]
\protect\caption{\label{tab:A-horse-race-ROBUST}A robust horse race of cross-validated
estimations.}

\centering{}\includegraphics[width=1\textwidth]{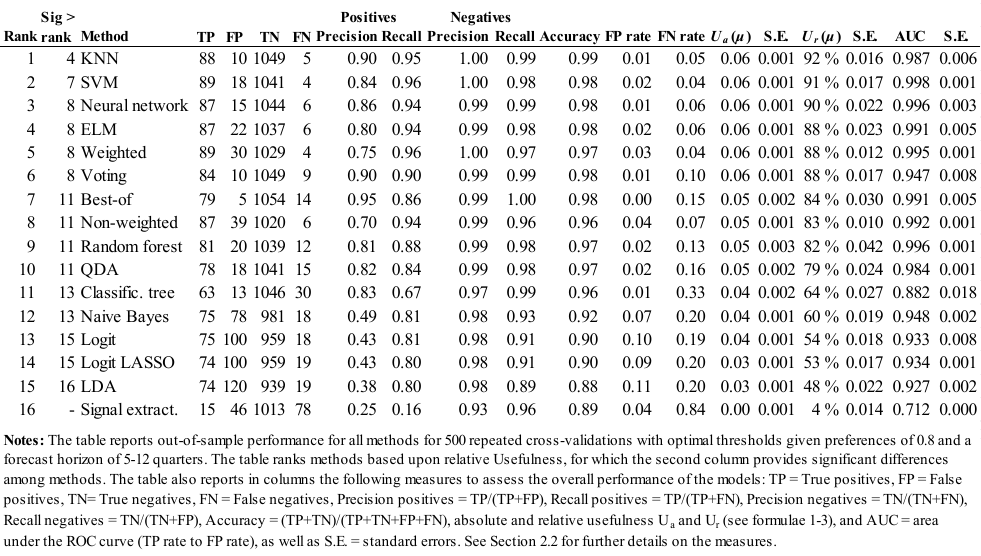}
\end{table}

To again perform the more stringent recursive real-time evaluation,
but as a robust exercise, we combine the recursive horse race with
double resampling. In Table \ref{tab:Robust recursive comparison},
we draw 500 bootstrap samples of in-sample data for each quarter,
and again report average out-of-sample results, including its SE.
In comparison with the results for the single estimations in Table
\ref{tab:Recursive comparison}, the rankings exhibit slight differences.
Whilst most machine learning methods still outperform the more conventional
methods, the difference is smaller in general. In particular, ANN
exhibits best Usefulness among the individual methods, while its counterpart
SVM performs worse than in the single estimations. Most notably, Logit
LASSO and classification trees show a positive increase in ranking.
Again, based upon the statistical significances of the cross-comparison
matrix in Table \ref{tab:Significance_rec} in the Appendix, we report
significant differences in ranks in the second column of Table \ref{tab:Robust recursive comparison}.
Compared to the cross-validation exercise, the variation in in-sample
data introduced by the double bootstrap has a notable effect on the
variation in performance, and hence also on the significant differences
in ranks. The top three methods in Table \ref{tab:Robust recursive comparison}
are aggregates, being the only methods statistically significantly
better than any other method than signal extraction. Next is a large
intermediate group of approaches, with signal extraction being the
worst-in-class method. Again, we also provide recursive out-of-sample
ROC curve plots for all methods and the aggregates in Figure \ref{fig:AUC-rec}
in the Appendix.

In line with this, as there is no one single performance measure,
we also rank methods in both of the two exercises based upon their
AUC, compute their variation in the exercise and conduct equality
tests. For both the cross-validated and the recursive exercises, these
tables show coinciding results with the Usefulness-based rankings,
as is shown in the Appendix in \ref{tab:Significance_CV-AUC} and
\ref{tab:Significance_rec-AUC}. For cross-validated evaluations,
one key difference is that the AUC ranking shows better relative performance
for the random forest and the best-of and non-weighted aggregates,
whereas the KNN and QDA improve their ranking in the recursive exercise.

\begin{table}[H]
\protect\caption{\label{tab:Robust recursive comparison}A robust horse race of recursive
real-time estimations.}

\centering{}\includegraphics[width=1\textwidth]{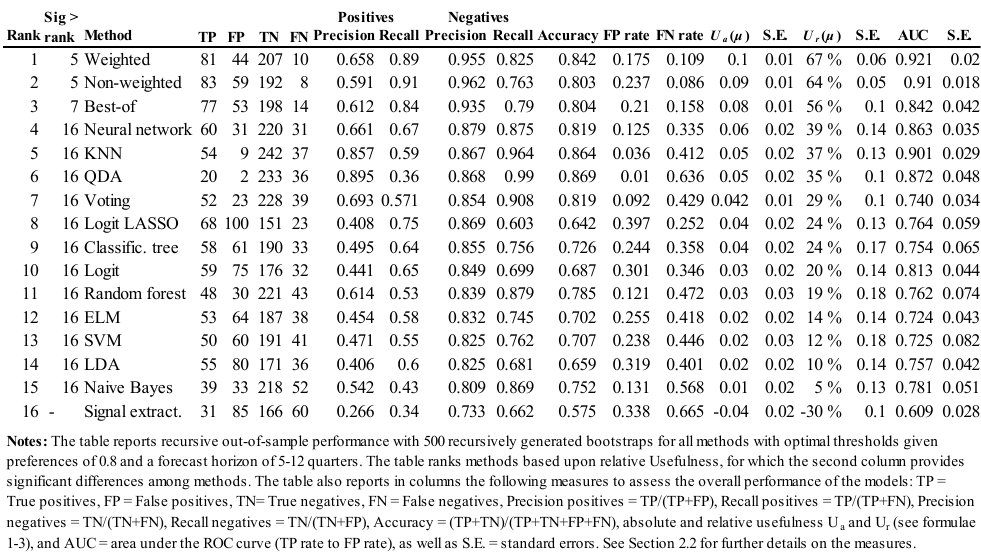}
\end{table}

\paragraph{Model output uncertainty}

\begin{figure}[H]
\begin{centering}
\includegraphics[width=0.52\textwidth]{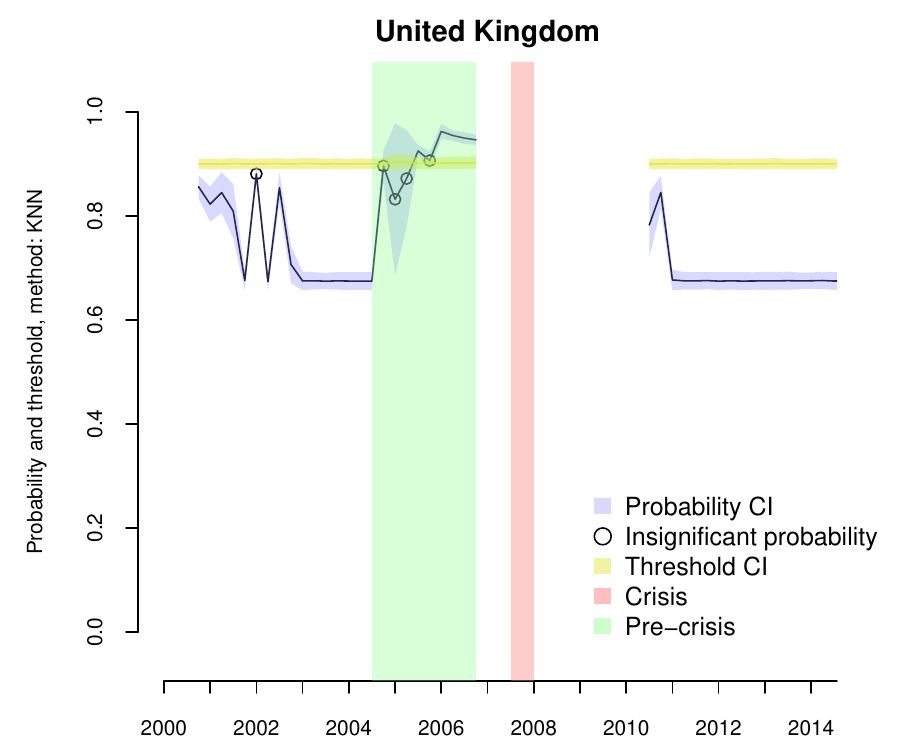}\includegraphics[width=0.52\textwidth]{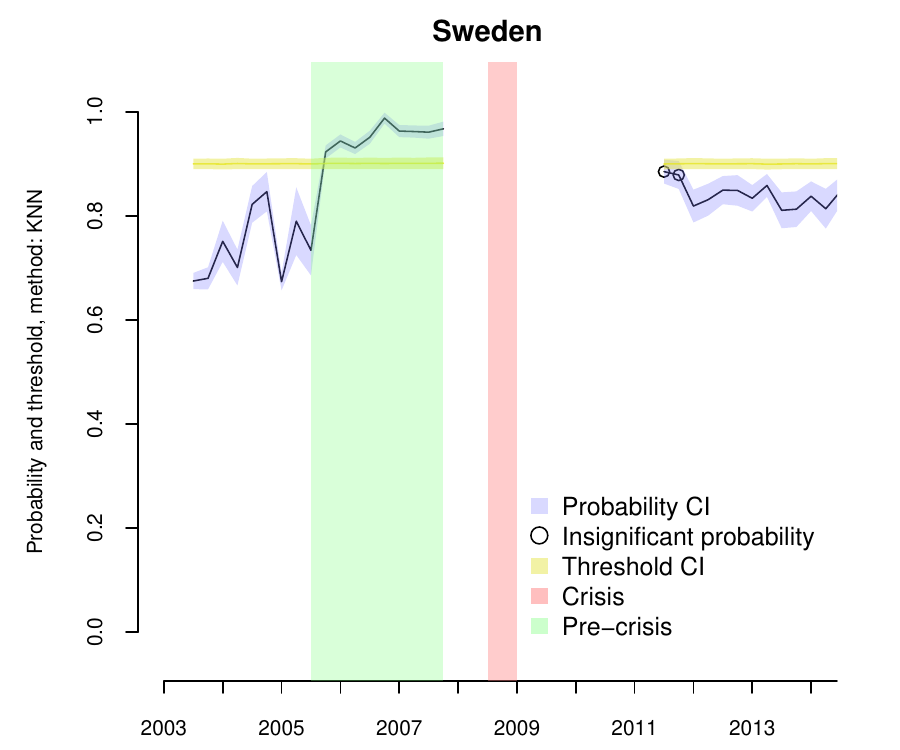}
\par\end{centering}

\centering{}\protect\caption{\label{fig:Confidence intervals- UK_SWE}Probabilities and thresholds,
and their CIs, of KNN for United Kingdom and Sweden}
\end{figure}

This section goes beyond pure measurement of classification performance
by first illustrating more qualitatively the value of representing
uncertainty for early-warning models. In line with Section \ref{sub:Model-uncertainty},
we provide confidence intervals (CIs) as an estimate of the uncertainty
in a crisis probability and its threshold. When computed for the aggregates,
we also capture increases in the variance of sample probabilities
due to disagreement in model output among methods, beyond variation
caused by resampling. In Figure \ref{fig:Confidence intervals- UK_SWE},
we show line charts with crisis probabilities and thresholds for United
Kingdom and Sweden from 2004Q1--2014Q1 for one individual method (KNN),
where tubes around lines represent CIs. The probability observations
that are not found to statistically significantly differ from a threshold
(i.e., above or below) are shown with circles. This represents uncertainty,
and hence points to the need for further scrutiny, rather than a mechanical
classification into vulnerable or tranquil periods. Thus, the interpretation
may indicate vulnerability for an observation to be below the threshold
or vice versa.

For UK, the left chart in Figure \ref{fig:Confidence intervals- UK_SWE}
illustrates first one elevated signal (but no threshold exceedance)
already in 2002, and then during the pre-crisis period larger variation
in elevated probabilities, which cause an insignificant difference
to the threshold and hence an indication of potential vulnerability.
This would have indicated vulnerability four quarters earlier than
without considering uncertainty. On the other hand, the right chart
in Figure \ref{fig:Confidence intervals- UK_SWE} shows for Sweden
that the two observations after a post-crisis period are elevated
but below the threshold. In the correct context, and in conjunction
with expert judgment, this would most likely not be related to a boom-bust
type of an imbalance, but rather elevated values in the aftermath
of a crisis.

As a next step in showing the usefulness of incorporating uncertainty
in models, we conduct an early-warning exercise in which we disregard
observations whose probabilities $p_{n}^{m}$ and $p_{n}^{a}$ do
not statistically significantly differ from thresholds $\tau_{n}^{m}$
and $\tau_{n}^{a}$, respectively. Due to larger data needs in the
recursive exercise, which would leave us with small samples, we only
conduct a cross-validated horse race of methods, as well as compare
it to the exercise in Table \ref{tab:A-horse-race-ROBUST}. In this
case, the cross-validated exercise functions well as a test of the
impact of disregarding insignificant observations on early-warning
performance. In Table \ref{tab:A-horse-race-ROBUSTSIG}, rather than
focusing on specific rankings of methods, we enable a comparison of
the results of the new performance evaluation to the full results
in Table \ref{tab:A-horse-race-ROBUST}.\footnote{Voting is not considered as there is no direct approach to deriving
statistical significance of binary majority votes.} With the exception of signal extraction, which anyhow exhibits low
Usefulness, we can observe that all methods yield better performance
when dropping insignificant observations. While this is intuitive,
as the dropped observations are borderline cases, the results mainly
function as general-purpose evidence of our model output uncertainty
measure and the usefulness of considering statistical significance
vis-à-vis thresholds.

\begin{table}[H]
\protect\caption{\label{tab:A-horse-race-ROBUSTSIG}A robust and significant horse
race of cross-validated estimations.}

\centering{}\includegraphics[width=1\textwidth]{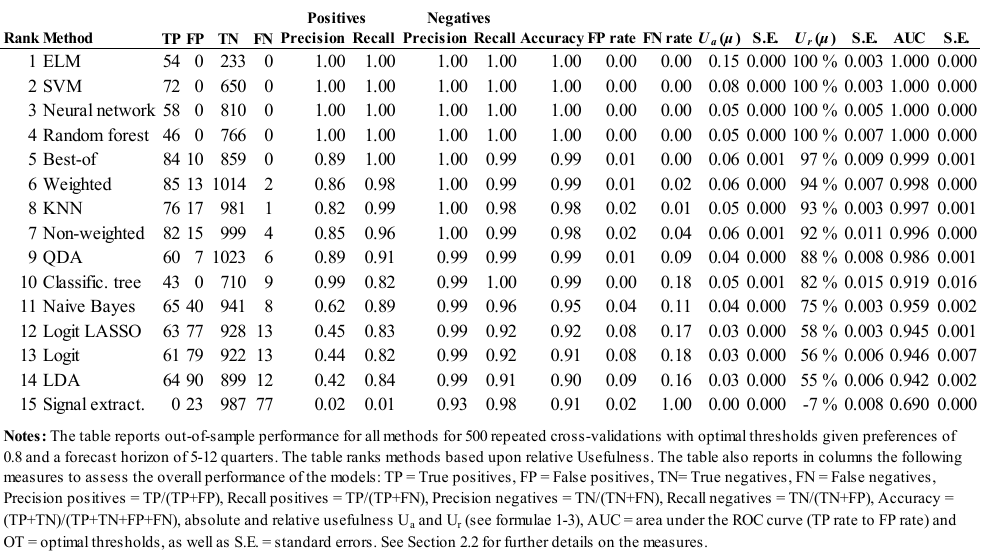}
\end{table}

\section{Conclusion}

This paper has presented first steps toward robust early-warning models.
As early-warning models are oftentimes built in isolation of other
methods, the exercise is of high relevance for assessing the relative
performance of a wide variety of methods. 

We have conducted a cross-validated and recursive horse race of conventional
statistical and more recent machine learning methods. This provided
information on best-performing approaches, as well as an overall ranking
of early-warning methods. The value of the horse race descends from
its robustness and objectivity. Further, we have tested four structured
approaches to aggregating the information products of built early-warning
models. Two structured approaches involve choosing the best method
(in-sample) for out-of-sample use, and relying on the majority vote
of all methods together. Then, moving toward more standard ensemble
methods for the use of multiple modeling techniques, we combined model
outputs into an arithmetic mean and performance-weighted mean of all
methods. Finally, we provided approaches for estimating model uncertainty
in early-warning exercises. One approach to tackling model performance
uncertainty, and provide robust rankings of methods, is the use of
mean-comparison tests on model performance. Also, we allow for testing
whether differences among the model output and thresholds are statistically
significantly different, as well as show that accounting for this
in signaling exercises yields added value. All approaches put forward
in this paper have been applied in a European setting, particularly
in predicting the still ongoing financial crisis using a broad set
of indicators. Generally, our results show that the conventional statistical
approaches are outperformed by more advanced machine learning methods,
such as $k$-nearest neighbors and neural networks, and particularly
by model aggregation approaches through ensemble learning.

The value and implications of this paper are manifold. First, we provide
an approach for conducting robust and objective horse races, as well
as an application to Europe. In relation to previous efforts, this
provides the first objective comparison of model performance, as we
assure a similar setting for each method when being evaluated, including
data, forecast horizons, post-crisis bias, loss function, policymaker's
preferences and overall exercise implementation. The robustness descends
from the use of resampling to assess performance, which assures stable
results not only with respect to small variation in data but also
for the non-deterministic modeling techniques. In the recursive real-time
exercises that control for non-linear function approximators overfitting
data, we still find recent machine learning approaches to outperform
conventional statistical methods. Beyond showing that machine learning
approaches have potential in these types of exercises, this also points
to the importance of using appropriate resampling techniques, such
as accounting for time dependence. Second, given the number of different
available methods, the use of multiple modeling techniques is a necessity
in order to collect information of different types of vulnerabilities.
This might involve the simultaneous use of multiple models in parallel
or some type of aggregation. In addition to improvements in performance
and robustness, this may be valuable due to the fact that some of
the more recent machine learning techniques are oftentimes seen as
opaque in their functioning and less interpretable. For instance,
if a majority vote of a panel of models points to a vulnerability,
preferences against one individual modeling approach are less of a
concern. Thus, as the ensemble models both perform well in horse races
and decrease variability in model performance, structured approaches
to aggregate model output ought to be one part of a robust early-warning
toolbox. Third, even though techniques and data for early-warning
analysis are advancing, and so is performance, it is of central importance
to understand the uncertainty in models. A key topic is to assure
that breaching a threshold is not due to sampling error alone. Likewise,
we should be concerned with observations below but close to a threshold,
particularly when the difference is not of significant size.

For the future, we hope that a large number of approaches for measuring
systemic risk, including those presented herein, are to be implemented
in a more structured and user-friendly manner. In particular, a broad
palette of measurement techniques requires a common platform for modeling
systemic risk and visualizing information products, as well as means
to interact with both model parameters and visual interfaces. This
could, for instance, involve the use of visualization and interaction
techniques provided in the VisRisk platform for visual systemic risk
analytics \citep{Sarlin2013SWIFT}, as well as more advanced data
and dimension reduction techniques \citep{Sarlin2013a,SarlinPRL2013}.
In conjunction with these types of interfaces, we hope that this paper
generally stimulates the simultaneous use of a broad panel of methods,
and their aggregates, as well as accounting for uncertainty when interpreting
results.

\newpage{}

\section*{\textmd{\small{}\renewcommand\refname{References}}}

{\small{}\bibliographystyle{plainnat}
\bibliography{References/references}
}{\small \par}

\section*{\textmd{\small{}\newpage{}}}

\setcounter{table}{0}

\setcounter{figure}{0}

\renewcommand{\thetable}{A.\arabic{table}}

\setcounter{section}{0}

\renewcommand{\thesection}{Appendix A}

\renewcommand{\thefigure}{A.\arabic{figure}}

\section{Robustness tests and additional results}

\begin{table}[H]
\protect\caption{\label{tab:Signalextraction}Cross-validated results for signal extraction.}

\centering{}\includegraphics[width=1\textwidth]{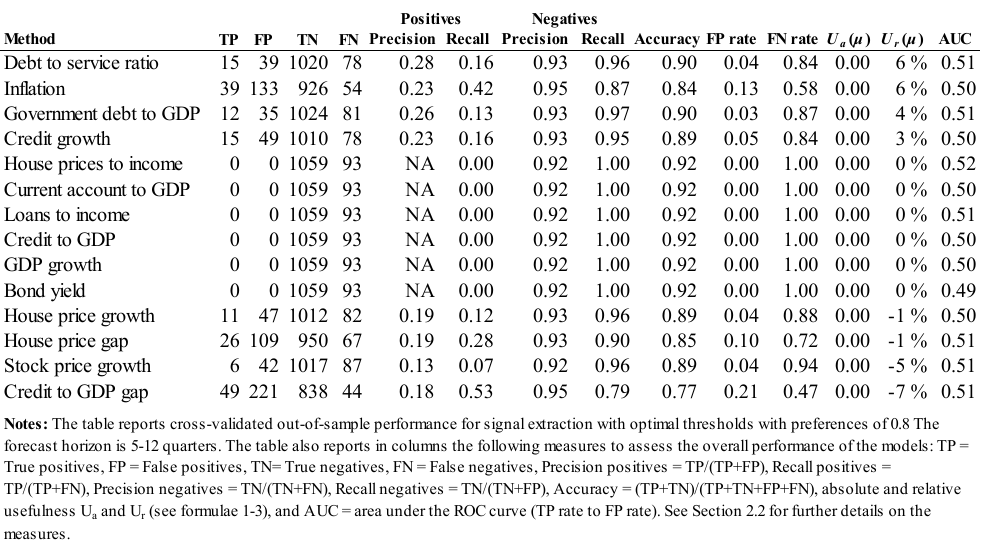}
\end{table}
\begin{table}[H]
\protect\caption{\label{tab:Signalextraction-mu0.91}Cross-validated results for signal
extraction with $\mu=0.9193$.}

\centering{}\includegraphics[width=1\textwidth]{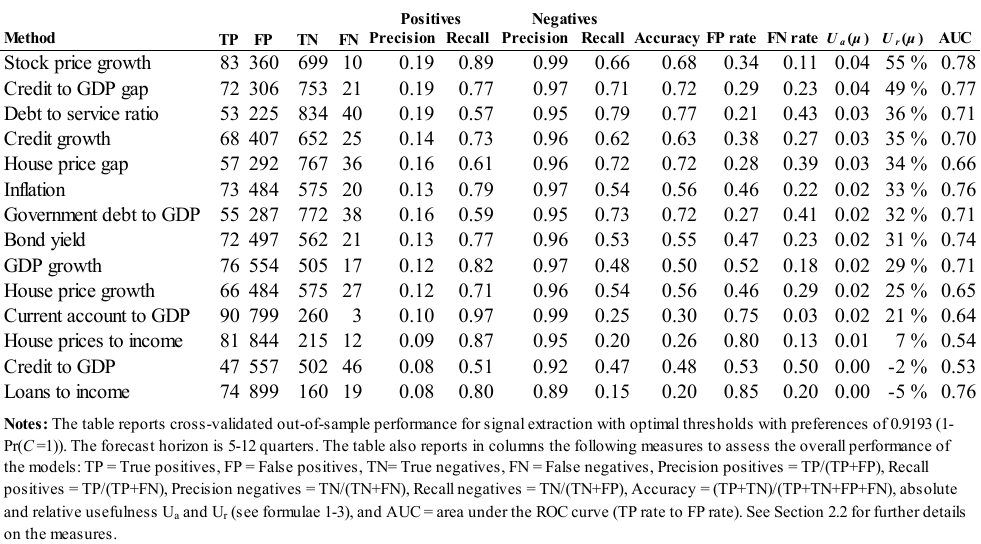}
\end{table}

\newpage{}

\begin{table}[H]
\protect\caption{\label{tab:Recursive comparison-ROB}A horse race of recursive real-time
estimations with dropped windows.}

\centering{}\includegraphics[width=1\textwidth]{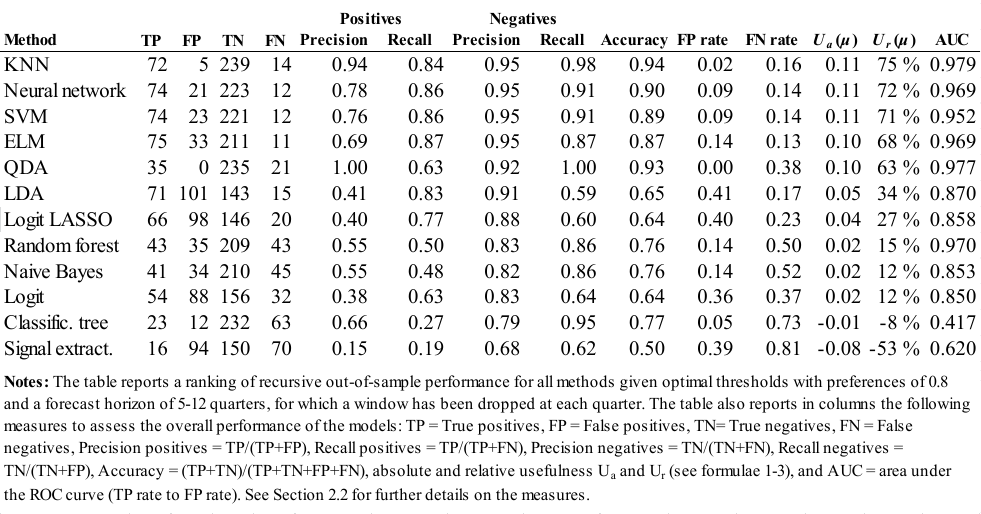}
\end{table}

\begin{table}[H]
\protect\caption{\label{tab:Aggregated-results-of-ROB}Aggregated results of recursive
estimations with dropped windows.}

\centering{}\includegraphics[width=1\textwidth]{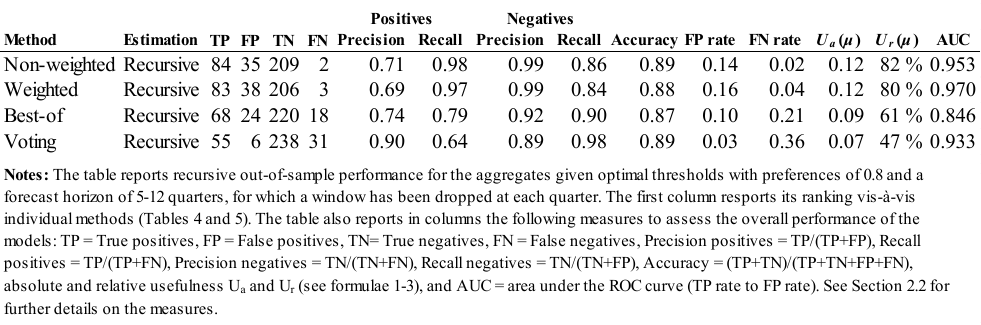}
\end{table}

\begin{table}[H]
\protect\caption{\label{tab:Significance_CV}Significances of cross-validated Usefulness
comparisons.}

\centering{}\includegraphics[width=1\textwidth]{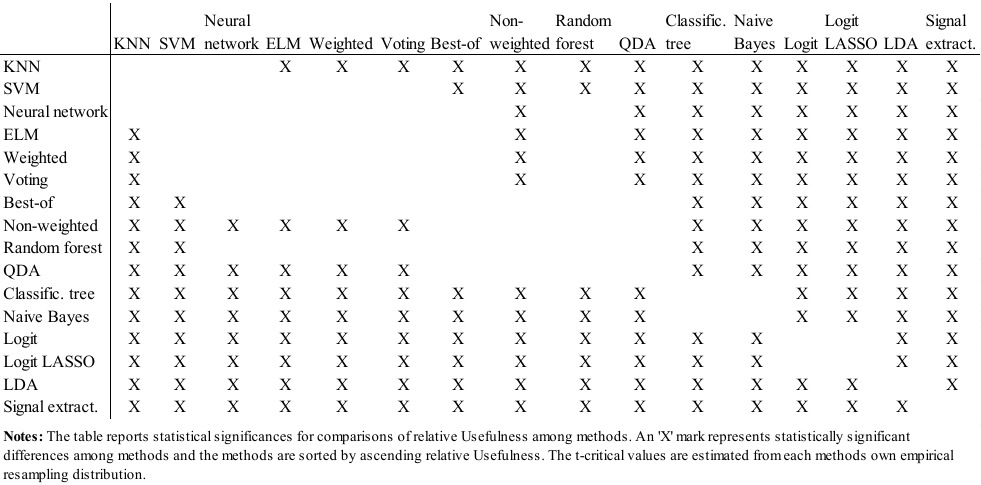}
\end{table}

\begin{table}[H]
\protect\caption{\label{tab:Significance_rec}Significances of recursive Usefulness
comparisons.}

\centering{}\includegraphics[width=1\textwidth]{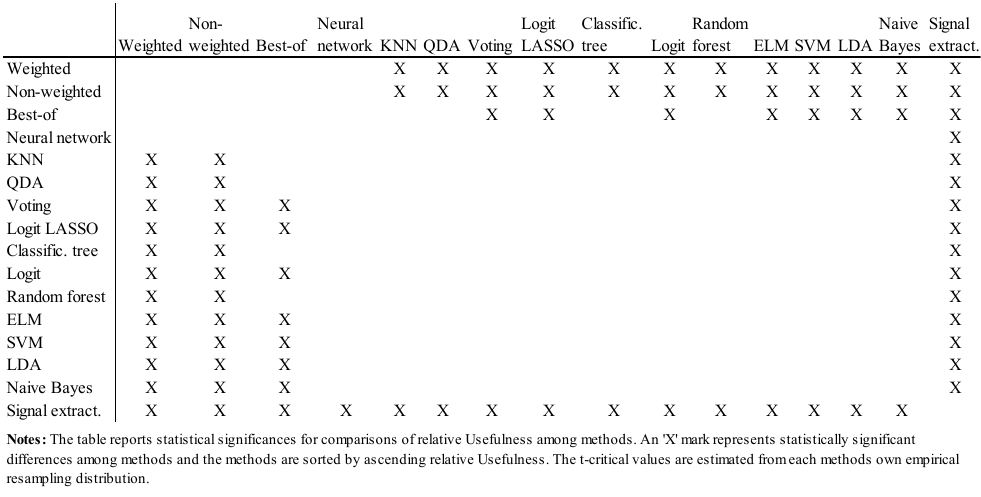}
\end{table}

\begin{table}[H]
\protect\caption{\label{tab:Significance_CV-AUC}Significances of cross-validated AUC
comparisons.}

\centering{}\includegraphics[width=1\textwidth]{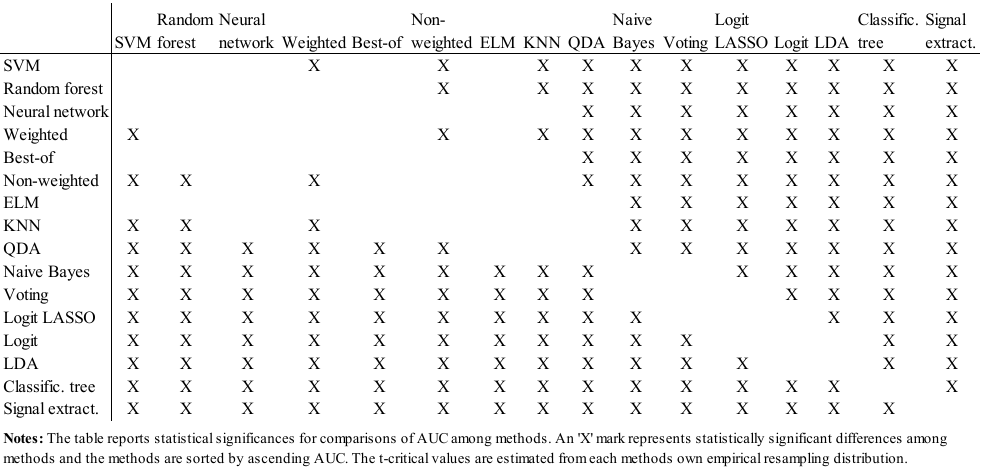}
\end{table}

\begin{table}[H]
\protect\caption{\label{tab:Significance_rec-AUC}Significances of recursive AUC comparisons.}

\centering{}\includegraphics[width=1\textwidth]{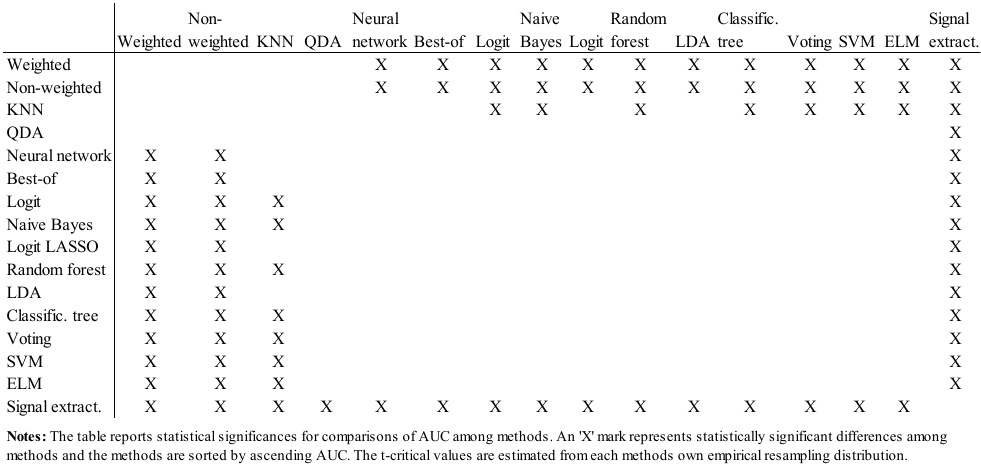}
\end{table}

\newpage{}

\begin{figure}[H]
\begin{centering}
\includegraphics[width=0.3\textwidth]{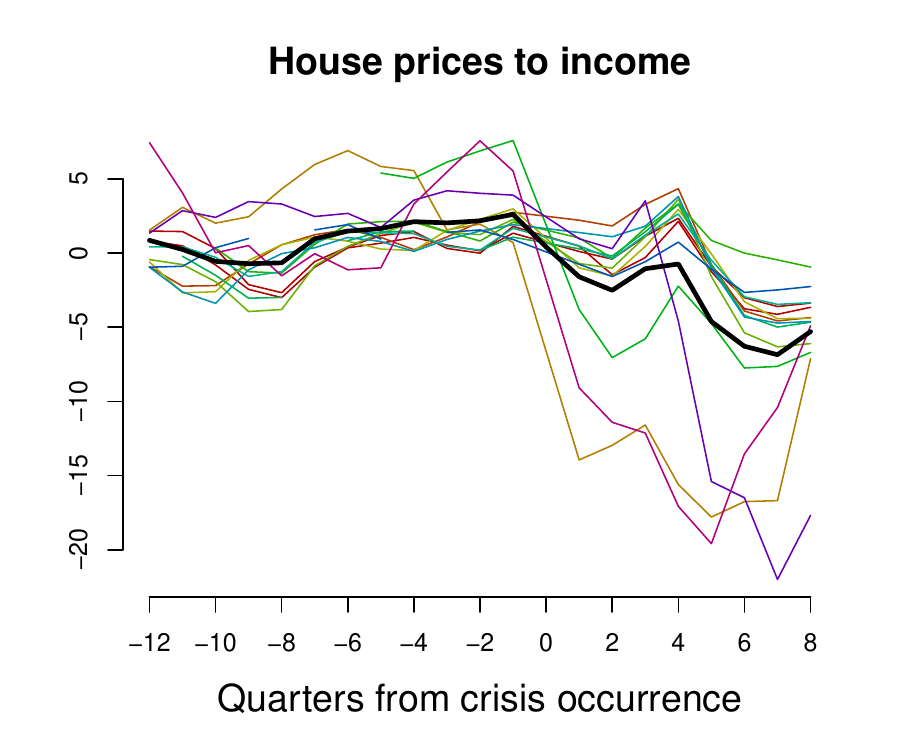}\includegraphics[width=0.3\textwidth]{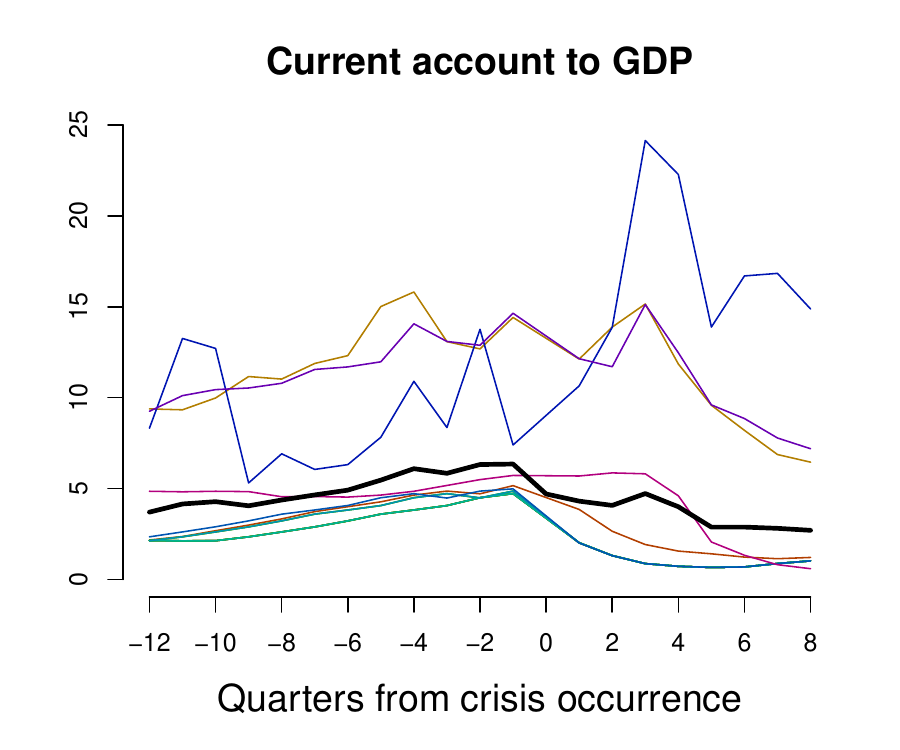}\includegraphics[width=0.3\textwidth]{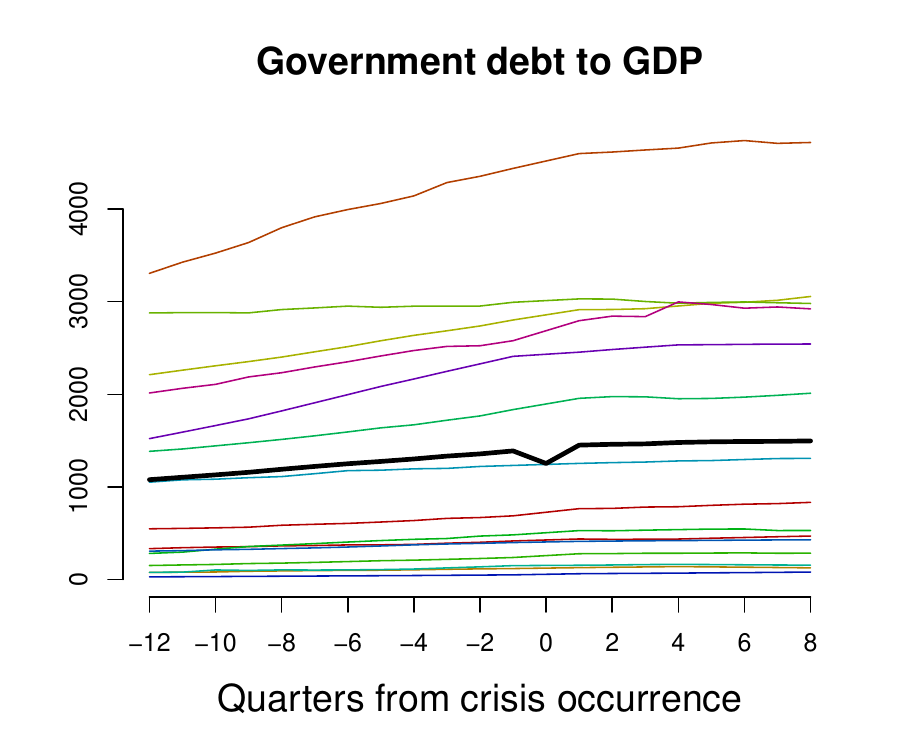}
\par\end{centering}

\begin{centering}
\includegraphics[width=0.3\textwidth]{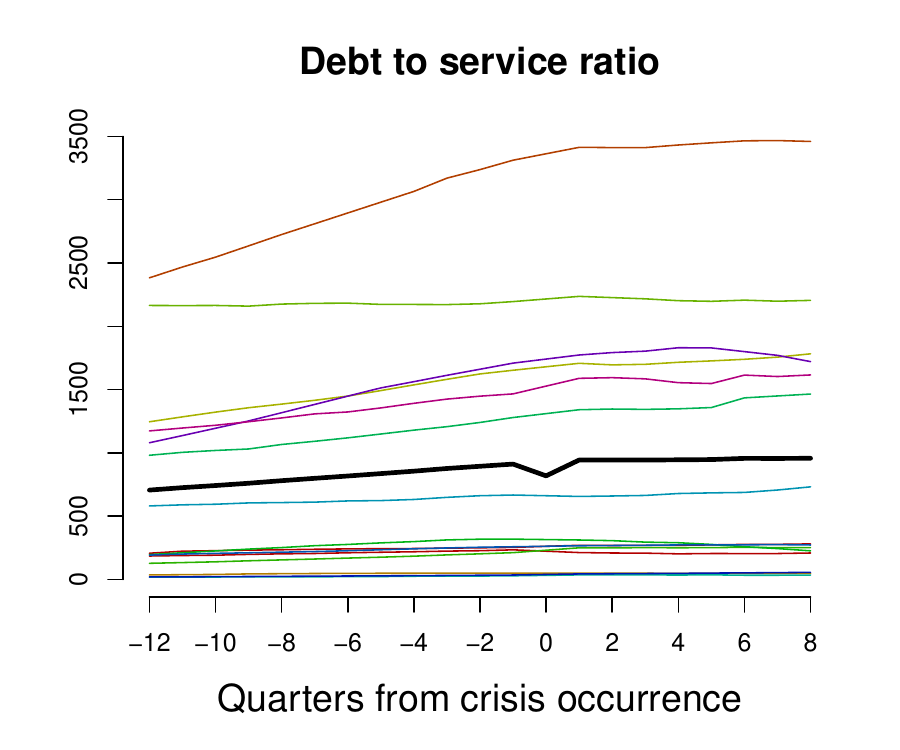}\includegraphics[width=0.3\textwidth]{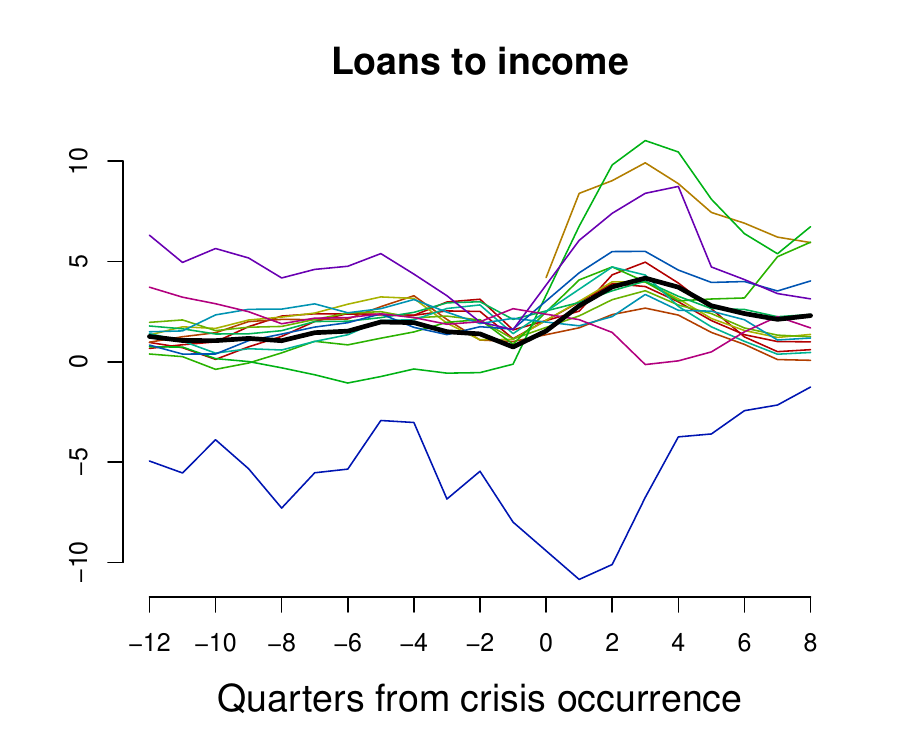}\includegraphics[width=0.3\textwidth]{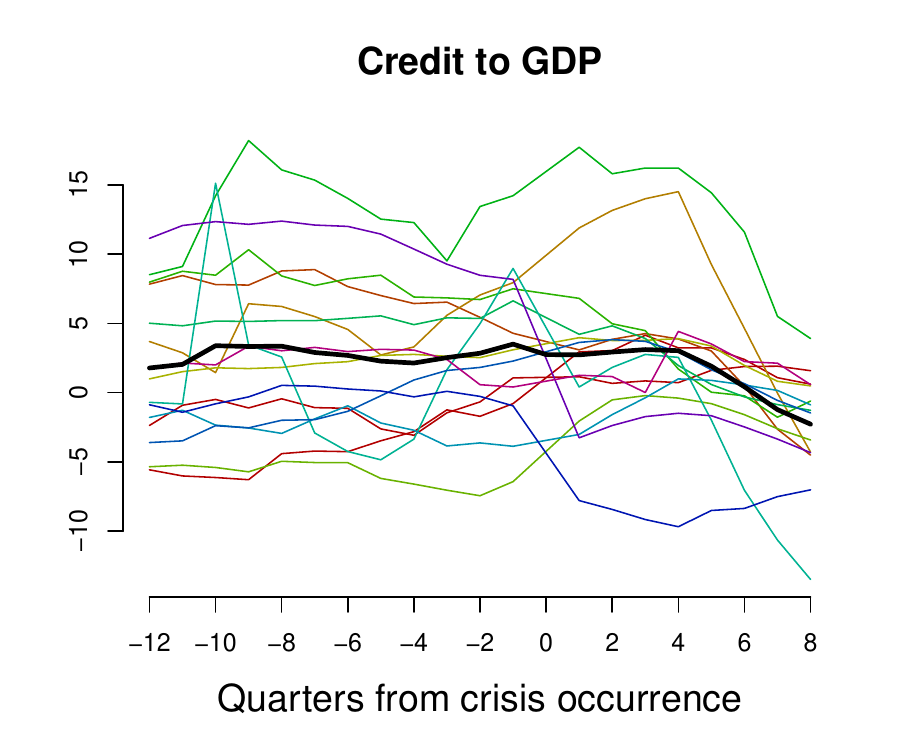}
\par\end{centering}

\begin{centering}
\includegraphics[width=0.3\textwidth]{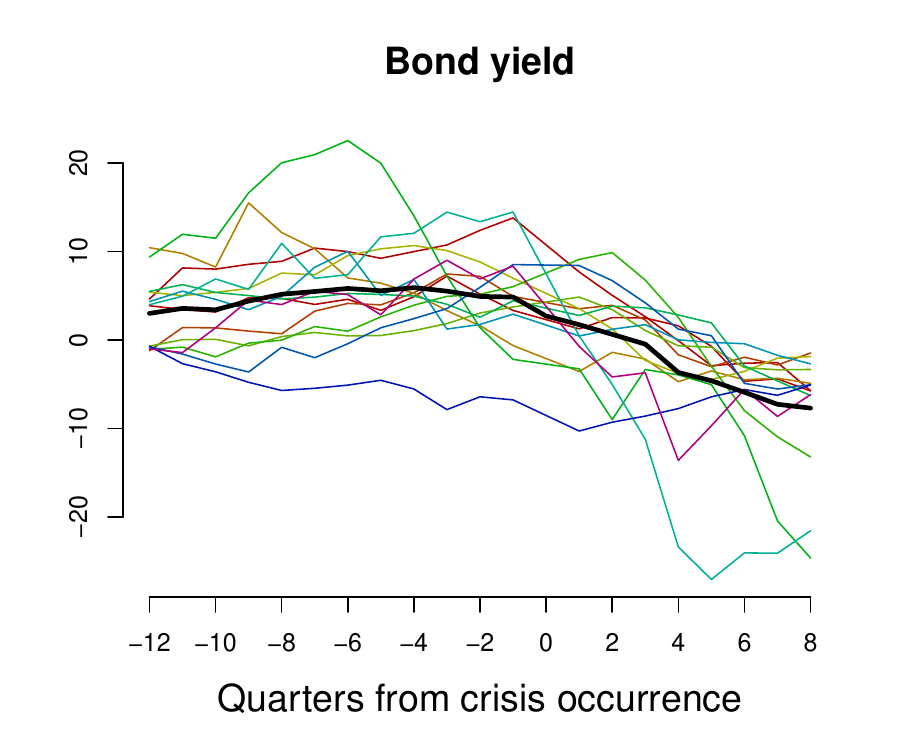}\includegraphics[width=0.3\textwidth]{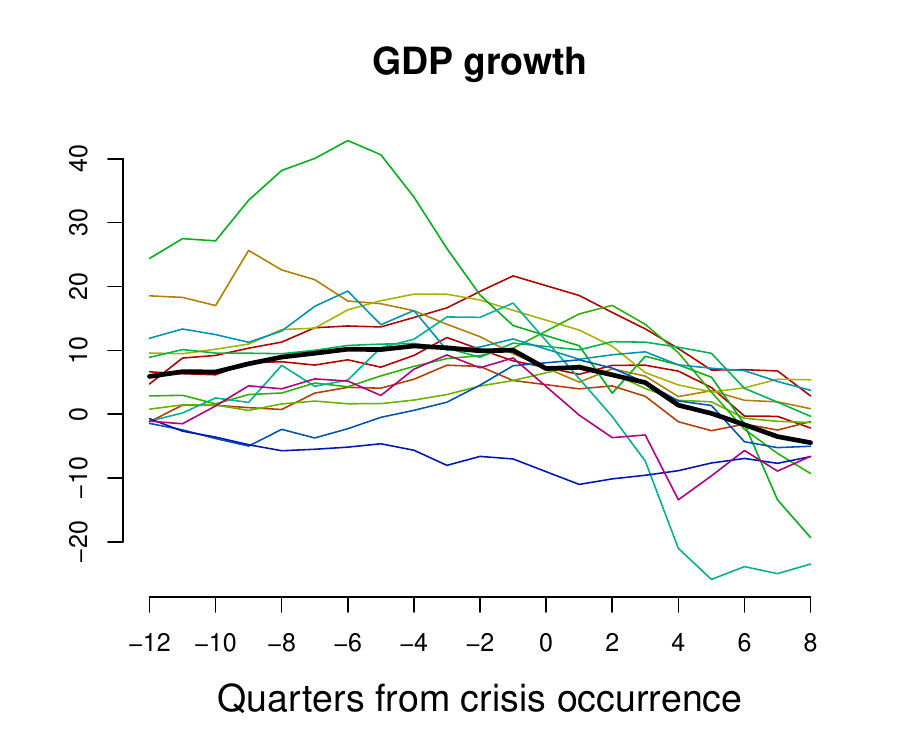}\includegraphics[width=0.3\textwidth]{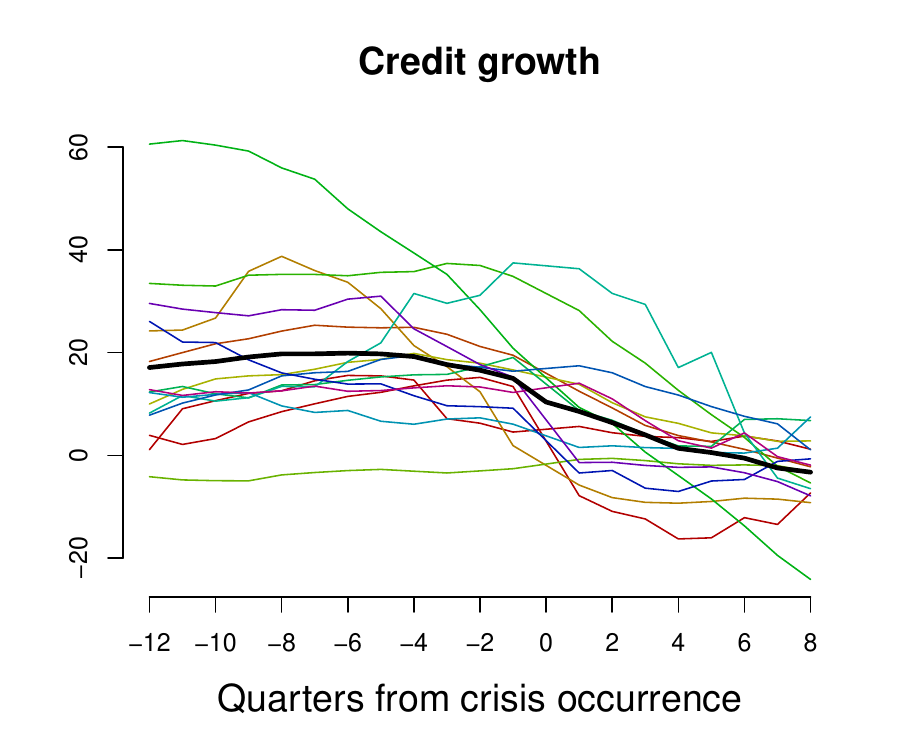}
\par\end{centering}

\begin{centering}
\includegraphics[width=0.3\textwidth]{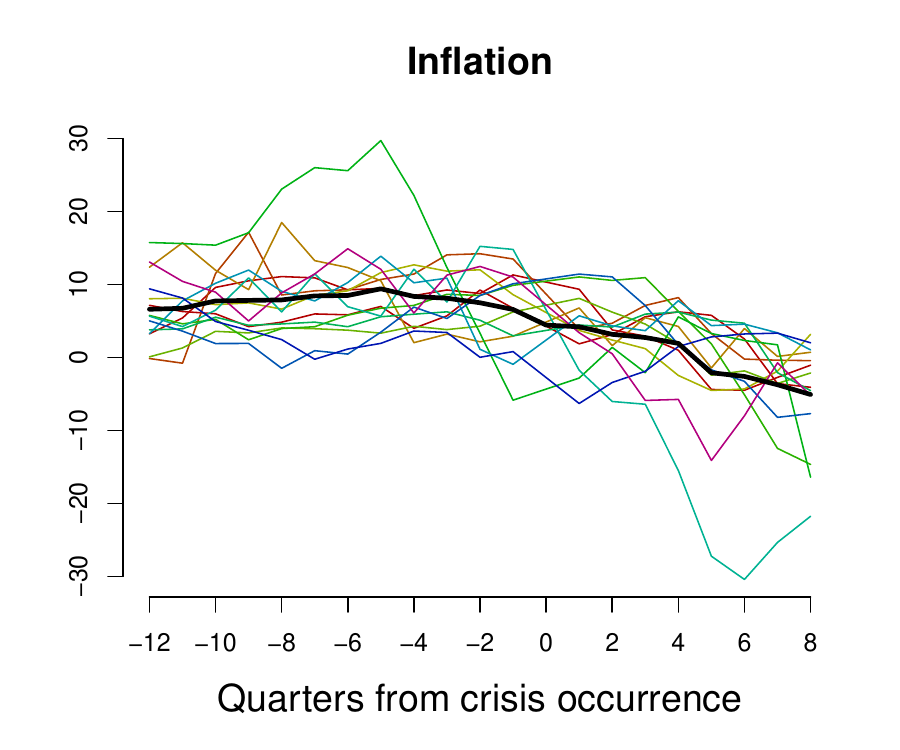}\includegraphics[width=0.3\textwidth]{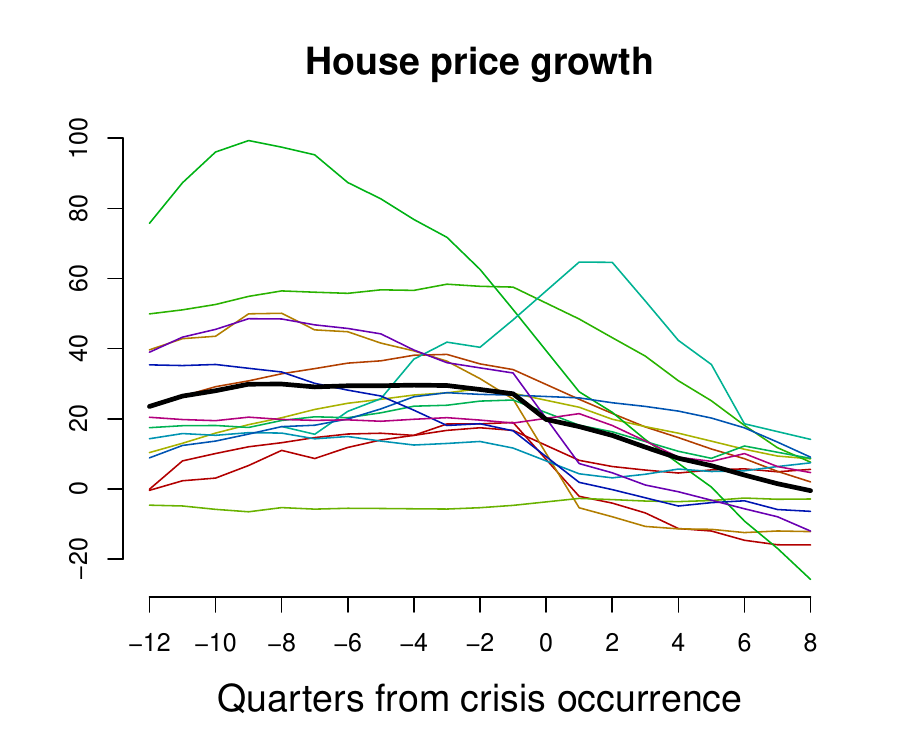}\includegraphics[width=0.3\textwidth]{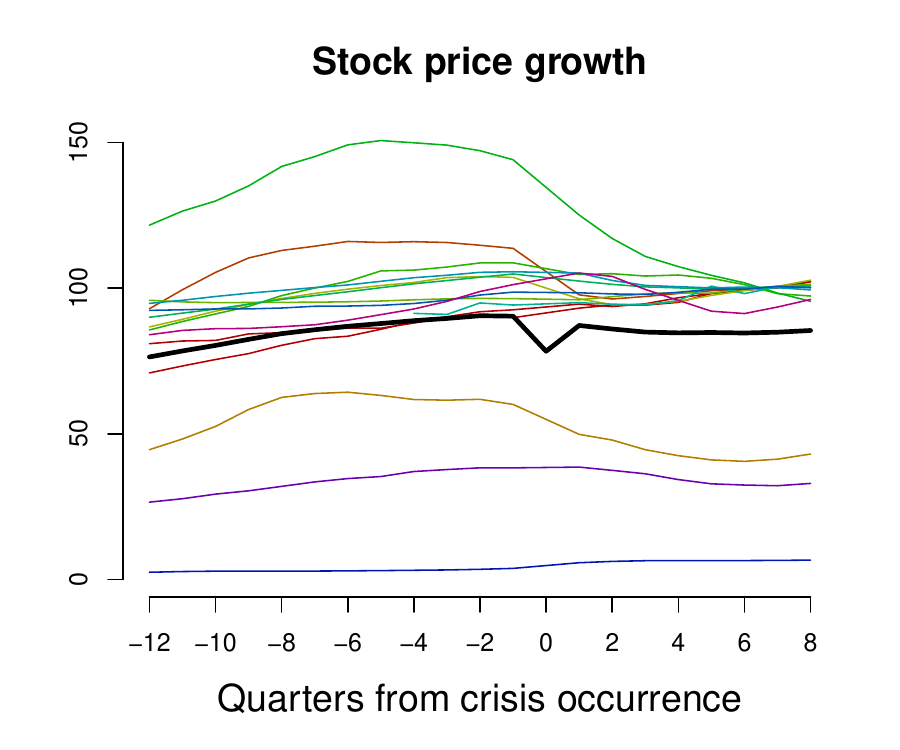}
\par\end{centering}

\begin{centering}
\includegraphics[width=0.3\textwidth]{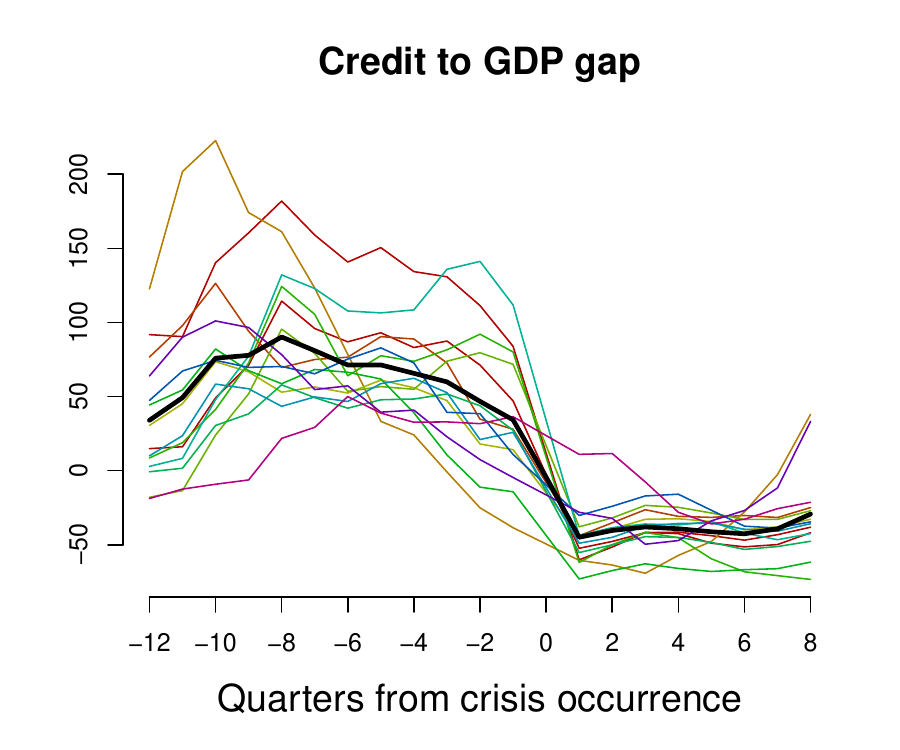}\includegraphics[width=0.3\textwidth]{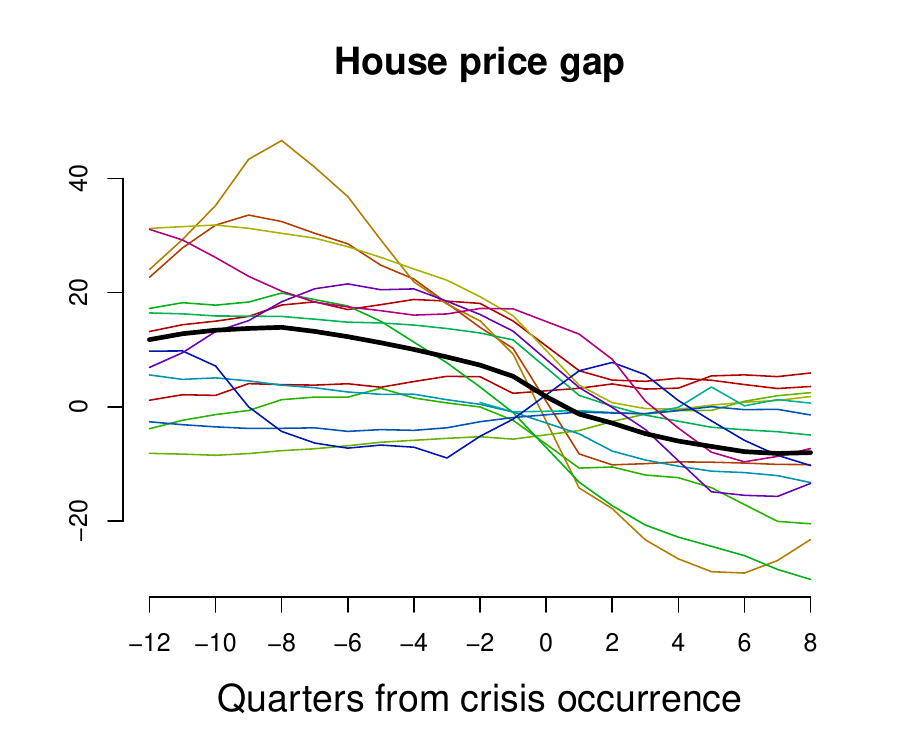}
\par\end{centering}

\protect\caption{\label{fig:Time-series-plots-of-indicators}Plots of each indicator
from $t-12$ to $t+8$ around crisis occurrences for each country.
The average of all entities is depicted as a bold line.}
\end{figure}

\begin{figure}[H]
\begin{centering}
\includegraphics[width=0.5\textwidth]{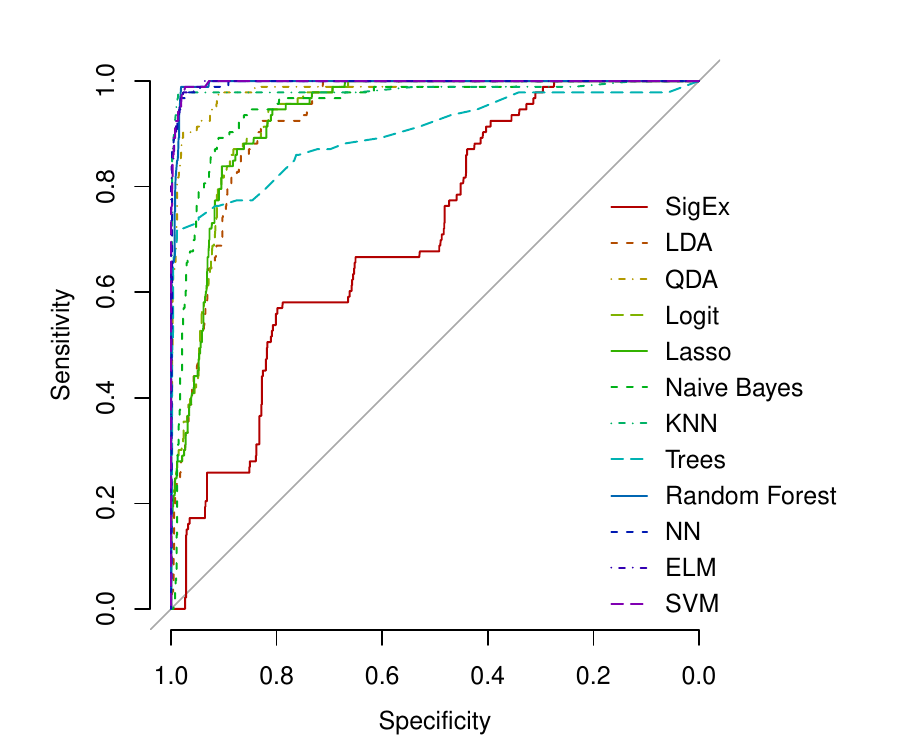}\includegraphics[width=0.5\textwidth]{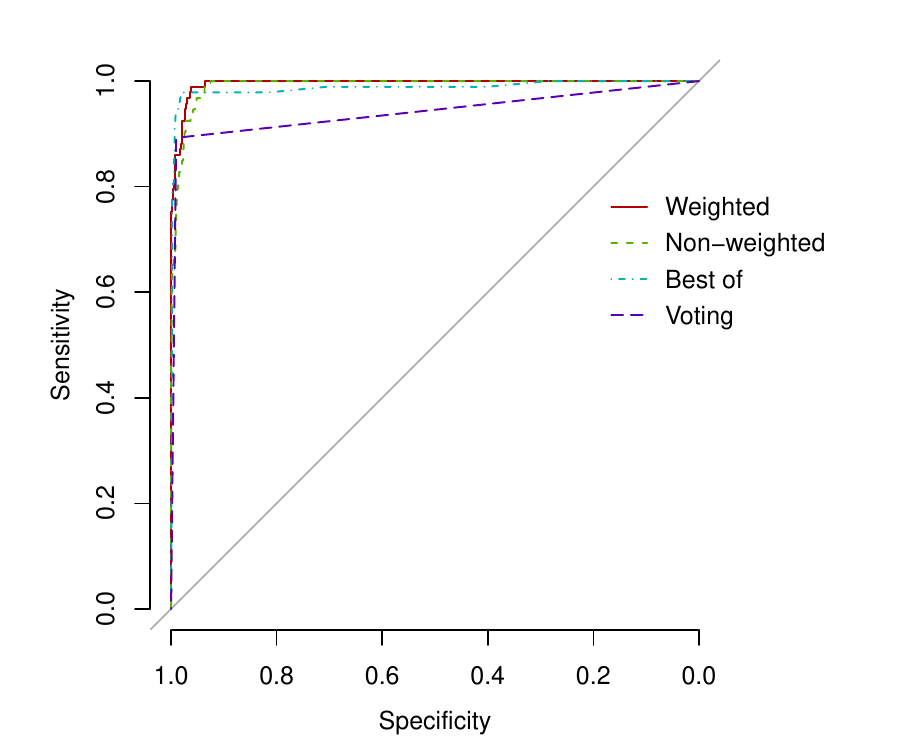}
\par\end{centering}

\centering{}\protect\caption{\label{fig:AUC-CV}Cross-validated out-of-sample ROC curve plots for
all methods and the aggregates}
\end{figure}

\begin{figure}[H]
\begin{centering}
\includegraphics[width=0.5\textwidth]{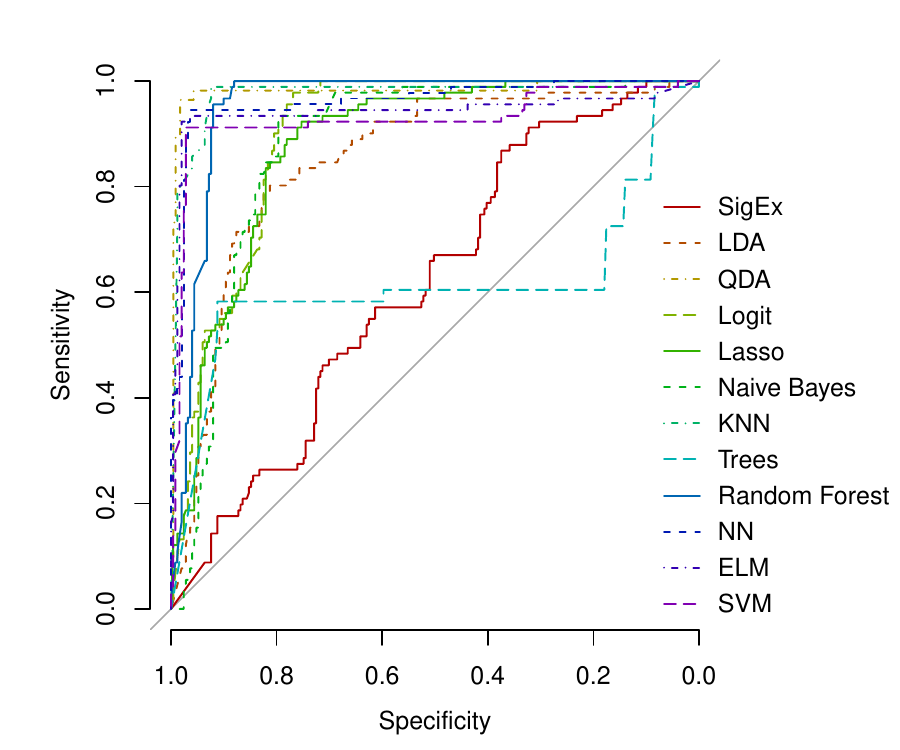}\includegraphics[width=0.5\textwidth]{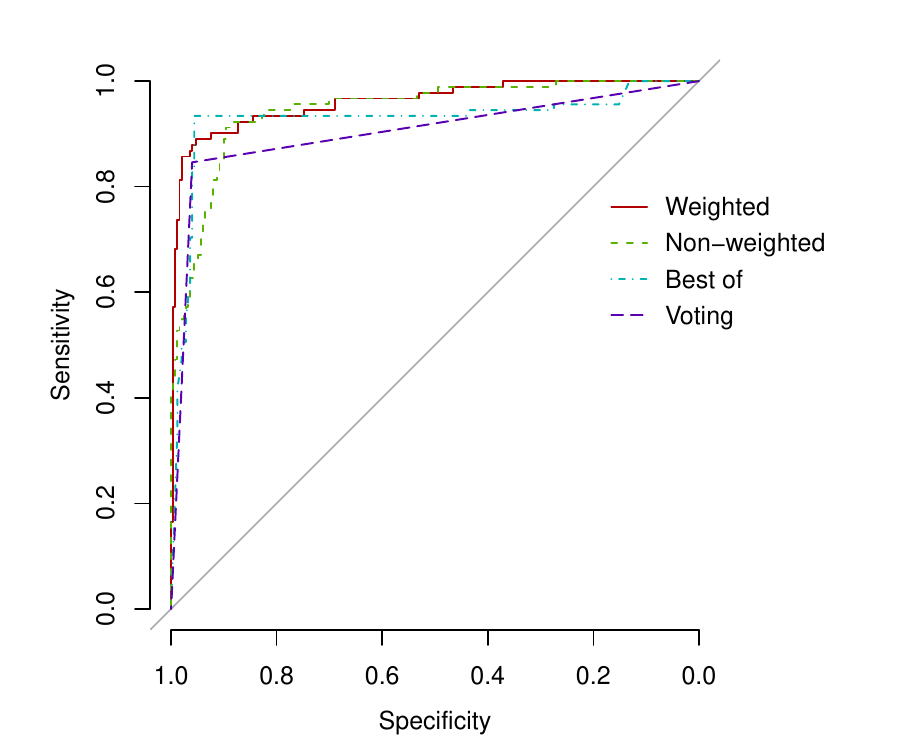}
\par\end{centering}

\centering{}\protect\caption{\label{fig:AUC-rec}Recursive out-of-sample ROC curve plots for all
methods and the aggregates}
\end{figure}

\end{document}